\documentclass[fleqn,10pt]{wlscirep}
\usepackage[utf8]{inputenc}
\usepackage[T1]{fontenc}
\usepackage{bm}
\usepackage{amsmath}
\usepackage{esint}
\usepackage{hyperref}
\usepackage{siunitx}
\usepackage{mhchem}
\usepackage{enumitem}
\usepackage[normalem]{ulem}
\usepackage{lineno}

\newcommand{\eq}[1]{Eq.~\eqref{#1}}
\newcommand{\eqs}[2]{Eqs.~\eqref{#1}-\eqref{#2}}
\newcommand{\fig}[1]{Fig.~\ref{#1}}
\newcommand{\sctn}[1]{\S~\ref{#1}}

\newcommand{\movie}[1]{\ref{#1}}

\newcommand{\FIG}[1]{Figure~\ref{#1}}

\newcommand{\figsi}[1]{Ext. Data Fig.~\ref{#1}}

\newcommand{\ie}{\textit{i.e.} }

\newcommand{\etal}{\textit{et.~al.}}

\newcommand{\av}[1]{\left\langle #1 \right\rangle}
\newcommand{\abs}[1]{\left| #1 \right|}
\newcommand{\tens}[1]{\boldsymbol{\bm{#1}}}
\newcommand{\grad}{\vec{\nabla}}
\newcommand{\laplace}{\nabla^2}
\renewcommand{\div}{\grad\cdot}
\newcommand{\curl}{\grad\times}
\newcommand{\transpose}[1]{{ #1 }^{T}}

\newcommand{\tr}[1]{\mathrm{tr}\left[ #1 \right] }

\newcommand{\sgn}[1]{\mathrm{sgn}\left[ #1 \right]}

\newcommand{\qcrit}{\mathcal{Q}}
\newcommand{\nemOrder}{Q}
\newcommand{\nemT}{\tens{\nemOrder}}
\newcommand{\molField}{H}
\newcommand{\molFieldT}{\tens{\molField}}
\newcommand{\dir}{n}
\newcommand{\dirV}{\vec{n}}
\newcommand{\vel}{u}
\newcommand{\velV}{\vec{\vel}}
\newcommand{\pos}{r}
\newcommand{\posV}{\vec{\pos}}
\renewcommand{\time}{t}
\newcommand{\spacetime}{\left(\posV,\time\right)}
\newcommand{\coRot}{R}
\newcommand{\vort}{\Omega}
\newcommand{\vortT}{\tens{\vort}}
\newcommand{\vortvec}{\omega}
\newcommand{\vortvecV}{\vec{\vortvec}}
\newcommand{\strainrate}{E}
\newcommand{\strainrateT}{\tens{\strainrate}}
\newcommand{\gradVel}{L}
\newcommand{\gradVelT}{\tens{\gradVel}}
\newcommand{\unitT}{\tens{\delta}}
\newcommand{\stress}{\Pi}
\newcommand{\stressT}{\tens{\stress}}
\newcommand{\disclinDens}{D}
\newcommand{\disclinDensT}{\tens{\disclinDens}}
\newcommand{\ssb}{S_{\text{SB}}}
\newcommand{\visc}{\eta}
\newcommand{\act}{\zeta}
\newcommand{\shapeani}{\kappa}
\newcommand{\friction}{\kappa}
\newcommand{\frictionT}{\tens{\friction}}
\newcommand{\dens}{\rho}
\newcommand{\actlen}{\ell_\zeta}
\newcommand{\defectdens}{\sigma}
\newcommand{\force}{f}

\newcommand{\actForce}{\force_{\act}}
\newcommand{\actForceV}{\vec{\force}_{\act}}
\newcommand{\green}{G}
\newcommand{\greenT}{\tens{\green}}

\DeclareSIUnit{\molar}{M}

\newcounter{siequation}
\newcounter{sifigure}
\newcounter{sisection}
\newcounter{simovie}
\definecolor{pumpkin}{rgb}{1.0,0.4,0.0}
\definecolor{midnight}{rgb}{0.003921569,0.098039216,0.576470588}
\definecolor{saphire}{rgb}{0.0,0.196,0.372549}
\definecolor{crimson}{rgb}{0.75686,0,0.262745}
\definecolor{capri}{rgb}{0.0,0.768627,0.8745098}
\definecolor{amber}{rgb}{0.95686,0.66666667,0.0}
\definecolor{plum}{rgb}{0.50588,0.007843,0.3843137}
\definecolor{cerulean}{rgb}{0.0,0.568627,0.70980}
\definecolor{ruby}{rgb}{0.83137,0.0,0.4470588}

\title{Spontaneous Self-Constraint in Active Nematic Flows}

\author[1]{Louise C. Head}
\author[2]{Claire Dor\'e}
\author[1]{Ryan Keogh}
\author[3]{Lasse Bonn}
\author[3]{Amin Doostmohammadi}
\author[3]{Kristian Thijssen}
\author[2]{Teresa L\'opez-Le\'on}
\author[1,*]{Tyler N. Shendruk}
\affil[1]{School of Physics and Astronomy, The University of Edinburgh, Peter Guthrie Tait Road, Edinburgh, EH9 3FD, UK.}
\affil[2]{Laboratoire Gulliver, UMRS 7083, ESPCI Paris, PSL Research University, 75005 Paris, France}
\affil[3]{Niels Bohr Institute, University of Copenhagen, Blegdamsvej 17, Copenhagen 2100, Denmark}

\affil[*]{e-mail: t.shendruk@ed.ac.uk}

\begin{abstract}
Active processes drive and guide biological dynamics across scales---from subcellular cytoskeletal remodelling, through tissue development in embryogenesis, to population-level bacterial colonies expansion. 
In each of these, biological functionality requires collective flows to occur while self-organized structures are protected; however, the mechanisms by which active flows can spontaneously constrain their dynamics to preserve structure have not previously been explained. 
By studying collective flows and defect dynamics in active nematic films, we demonstrate the existence of a self-constraint---a two-way, spontaneously arising relationship between activity-driven isosurfaces of flow boundaries and mesoscale nematic structures. 
Our results show that self-motile defects are tightly constrained to viscometric surfaces---contours along which vorticity and strain-rate balance. 
This in turn reveals that self-motile defects break mirror symmetry when they move along a single viscometric surface, in contrast with expectations. 
This is explained by an interdependence between viscometric surfaces and bend walls---elongated narrow kinks in the orientation field. 
Although we focus on extensile nematic films, numerical results show the constraint holds whenever activity leads to motile half-charge defects. 
This mesoscale cross-field self-constraint offers a new framework for tackling complex 3D active turbulence, designing dynamic control into biomimetic materials, and understanding how biological systems can employ active stress for dynamic self-organization. 
\end{abstract}

\begin{document}

\flushbottom
\maketitle
\thispagestyle{empty}

\section{Introduction}\label{sctn:intro}

Disorderly turbulent flow exists in many classes of fluids, and characterizing inertial turbulence~\cite{Klotz2022} remains an outstanding challenge due to chaotic flows across spatial and temporal scales~\cite{Duraisamy2019}.
However, the challenges are compounded in rheologically complex fluids because couplings between fields introduces additional nonlinearities. 
In elastic~\cite{Duraisamy2019}, granular~\cite{Brandt2022}, magnetohydrodynamic~\cite{Dong2022}, quantum~\cite{Skrbek2021} and liquid crystaline~\cite{Takeuchi2007} turbulence, velocity is strongly coupled to other fields, some of which possess their own topologies with associated defects~\cite{Hosking2021,polanco2021,mur2022}. 
There is growing evidence that many biological systems spontaneously exhibit turbulence-like disorderly flow states that are coupled to local orientation (nematic) fields~\cite{Balasubramaniam2022,shankar2022}.
In these systems, active turbulence~\cite{wensink2012meso} is accompanied by continual creation/annihilation of topologically protected defects in the orientational field, including self-propelled $+1/2$ defects in 2D~\cite{sanchez2012} and disclination lines in 3D~\cite{Duclos2020}. 

To simplify the complexities of active nematic flows, studies often prioritize one of the fields. 
Namely, since self-motile defects have been clearly identified as crucial to active turbulence, they are often viewed as controlling their own dynamical evolution, with only perturbations---either due to local deformations or neighboring defects---causing deviations from their \textit{ideal} trajectories~\cite{shankar2019hydrodynamics,angheluta2021role,ronning2022}. 
From this perspective, hydrodynamic flows follow directly from the governing defect configuration. 
The antithetical approach is to integrate the orientational field dynamics directly into the hydrodynamic stresses creating defect-free fluid dynamical models~\cite{Bratanov2015,Alert2020,mukherjee2023}. 
Such velocity-fixated studies focus on long-lived and spatially extended structures in the velocity/vorticity fields and attempt to identify scaling laws for the energy/enstrophy spectra~\cite{Alert2022}. 
Together, these two approaches have made substantial progress towards understanding active turbulence, but the crucial mesoscale bridge between them remains to be developed. 

In this article, we reveal that non-linear coupling between flow and orientational fields in active nematics leads to a strong,  two-way, spontaneous self-constraint. 
On the one hand, we report that self-motile topological defects are tightly constrained to specific flow boundaries, while, on the other hand, these surfaces are driven by mesoscale defect-associated nematic deformations. 
Specifically, our results demonstrate that self-motile $+1/2$ defects are found solely on specific isosurfaces of flow boundaries identified as \textit{viscometric surfaces} --- contours where vorticity and strain-rate balance. 
These surfaces dictate evolution of the self-motile topological defects, which align with and move along the contours since there is no deformation of the streamlines along these paths. We uncover how this causes the defects to break their ideally predicted mirror symmetry and be classified according to their handedness. 
However, the spontaneous self-constraint is a co-dependency and we find that the viscometric surface, in turn, follows the mesoscopic deformation network of the orientational field. 
While we focus on 2D extensile active nematic turbulence, our conclusions hold whenever activity leads to motile half-integer topological defects, such as in vortex lattices, flow-tumbling liquid-crystals, contractile active stress, friction-induced ordering, and 3D structures.
The generality of these novel perspectives may provide greater insight into a more universal understanding of fluidic systems with non-linear field couplings and topologically protected states, as well as suggest a mechanism by which biological morphology could be dynamically protected. 

\section{Executive summary of experiments and simulations}\label{sctn:briefMethods}

To explore the spontaneous self-constraint between disclinations and flow structures in active nematics, we employ a combination of experimental and numerical techniques. 
Experimentally, we create an active nematic film by self-assembling labelled filamentous microtubules bundles at oil-water interface with kinesin motor clusters~\cite{hardouin2022}, which act as cross-linkers and hydrolyse ATP suspended in the aqueous phase, fueling extensile active stresses in the microtubule film (\sctn{sctn:exp}). 
Microtubule bundles are imaged via confocal fluorescence microscopy, which allows both the nematic tensor field $\nemT$ to be inferred through coherence-enhanced diffusion filtering~\cite{ellis2018} and the velocity field $\velV$ with an optical flow method. 
Numerically, we simulate active nemato-hydrodynamics via a hybrid lattice Boltzmann approach~\cite{Rorai2022} (\sctn{sctn:sims}). 
We focus on 2D extensile nematics in steady-state active turbulence, which give rise to disorderly flow profiles and continuous defect creation/annihilation (\fig{fig:Figure_1} and \movie{mov:ExpAT}). 

To explore the interplay between topology and  flow, we calculate mesoscopic structures in each field. 
Defects are readily apparent (\fig{fig:Figure_1}), and their positions and orientation are found from the nematic field $\nemT$ (\sctn{sctn:defectMethods}). 
Likewise, active flow fields have mesoscale structure with well-studied vorticity scales, which are visualized using the $\qcrit$-criterion~\cite{giomi2015}
\begin{align}
    \qcrit &= \frac{1}{2}\left( \lVert\vortT\rVert^2 - \lVert\strainrateT\rVert^2 \right) ,
    \label{eq:Qcriterion}
\end{align}
where $\vortT$ and $\strainrateT$ are the vorticity and strain-rate tensors (\sctn{sctn:sims}).
Vorticity-dominated regions are identified by $\qcrit>0$, while $\qcrit<0$ in strain-rate-dominated regions (\sctn{sctn:QCrit}). 
This technique has been used to identify vortex structures in confluent cell layers~\cite{blanch2018turbulent}, but any quantitative interplay between disclination positions, the $\qcrit$-criterion, and director structure has yet to be fully understood. 

\section{Plus-half defects are constrained to viscometric surfaces}

\begin{figure*}[t!]
    \centering
    \includegraphics[width=\columnwidth]{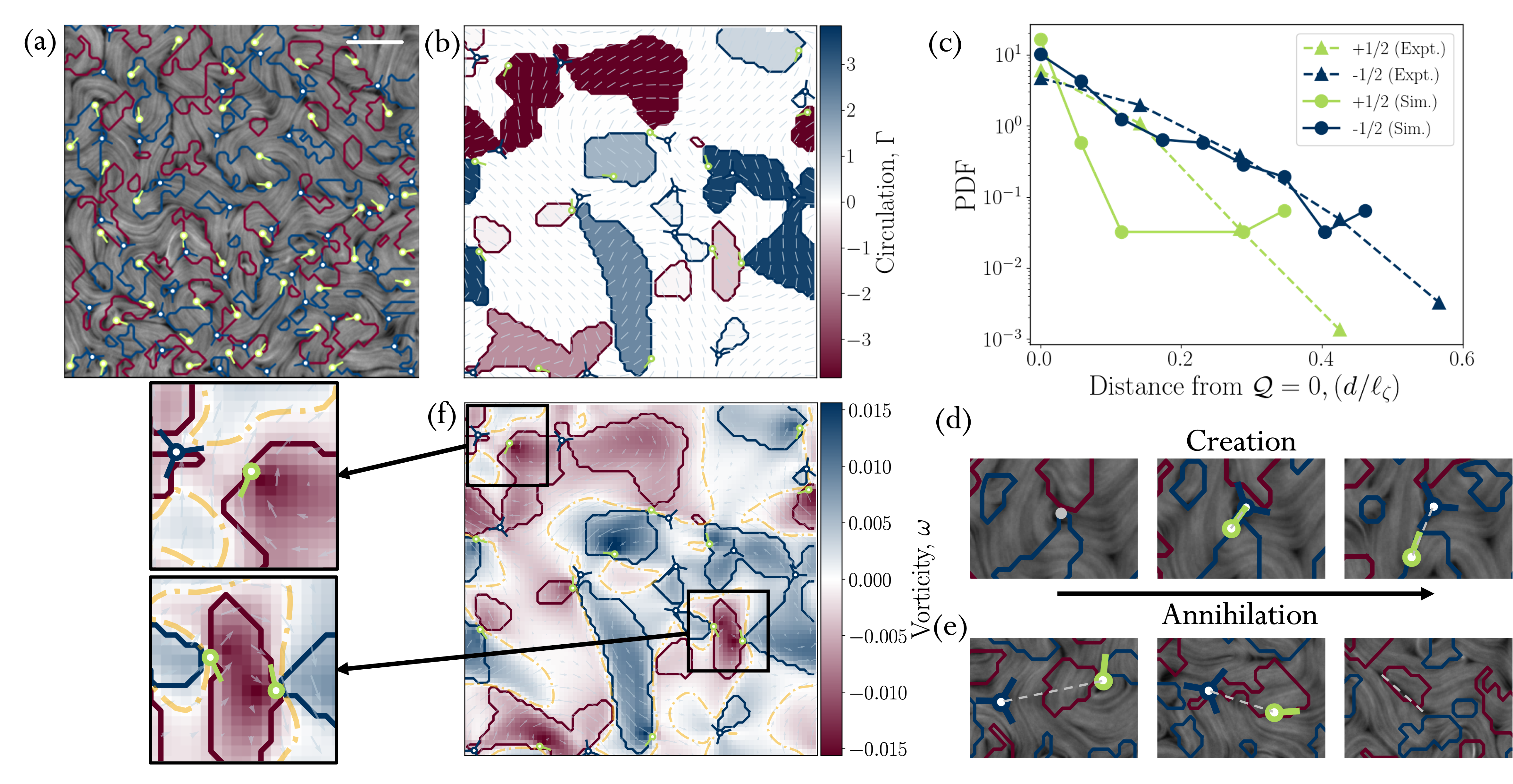}
    \caption{
    \textbf{Plus-half topological defects reside on isolines of $\qcrit=0$.} 
    \textbf{(a)} Scanning confocal microscopy of experimental active film (\sctn{sctn:exp}). 
    Plus-half defects marked as green comet-shaped symbols and minus-half defects by dark blue trefoil-shaped symbols (\sctn{sctn:defectMethods}). 
    Zero-isolines of $\qcrit$-criterion are shown as solid lines, colored by the handedness of the enclosed vortex---red for clockwise and blue for anti-clockwise (\sctn{sctn:QCrit}). Scale bar: \SI{100}{\micro\meter}. 
    \textbf{(b)} Numerical simulations of active turbulence (\sctn{sctn:sims}).  
    Circulation of each vortex colored representing the vortex strength. Nematic director field plotted as grey line field. 
    \textbf{(c)} Probability distribution function (PDF) of nearest distances of $\pm1/2$ defects to $\qcrit=0$ boundary (\sctn{sctn:interactions}).
    \textbf{(d)} Snapshots of pair creation event in  microtubule-kinesin film.
    \textbf{(e)} Experimental snapshots of pair annihilation event.
    \textbf{(f)} Same snapshot as (b), but vorticity field is colored, with grey arrows for the velocity field. 
    Zero-vorticity contours are shown as orange dash-dot lines (\sctn{sctn:vortStrainrateMag}). 
    Left: Examples where the vorticity is non-zero (top) and zero (bottom) at the $+1/2$ defect position. 
    } 
    \label{fig:Figure_1}%
\end{figure*}

By partitioning flow domains into vorticity-dominated ($\qcrit>0$) and strain-rate dominated ($\qcrit<0$) flow, we confirm the previously reported qualitative observation that defects tend to be found near the edge of vortices~\cite{giomi2015,shendruk2017dancing}. 
However, by imaging $\qcrit=0$ isolines (\sctn{sctn:QCrit}), we identify that $+1/2$ defects are always found precisely on the border where $\qcrit=0$ (\fig{fig:Figure_1}). 
We refer to the lines of $\qcrit=0$ as viscometric surfaces, as these are points where the velocity gradient tensor is singular and nilpotent indicating simple shear flow with no stretching or elongation of streamlines (\sctn{sctn:QCrit}). 
In the experimental system (\fig{fig:Figure_1}a; \movie{mov:ExpAT}), roughly one hundred defects are in frame and all of the $+1/2$ defects are located on viscometric surfaces (solid lines colored by the handedness of the enclosed vortex) at all times. 
Viscometric surfaces enclose vorticity-dominated regions, which exhibit a broad distribution of sizes, shapes and circulation (\figsi{fig:vortexstats}). 
Simulations reveal $\qcrit=0$ contours that have an associated $+1/2$ defect tend to enclose vortices with larger circulation (\fig{fig:Figure_1}b; \movie{mov:SimATCirculation}).
Both experiments and simulations demonstrate that, while $+1/2$ defects are constrained to viscometric surfaces, $-1/2$ defects are not (\fig{fig:Figure_1}c). 
Defect pair creation/annihilation events occur on $\qcrit=0$ boundaries (\fig{fig:Figure_1}d-e), with  the motile $+1/2$ defect pinned to the $\qcrit=0$ viscometric surface throughout the process, while the $-1/2$ defects are unbound from $\qcrit=0$ lines. 

Though previous studies observed that defects reside near the edges of vortices~\cite{giomi2015,shendruk2017dancing}, the implications have not been quantified. 
Commonly, active flow states are visualized by their vorticity field (\fig{fig:Figure_1}f; \movie{mov:SimATVelGradComp}). 
While these visualizations demonstrate the importance of vorticity in the characteristic active length and time scales, they are not observed to strictly constrain the configuration of defects. 
Identifing edges of vortices as the point where the vorticity $\vortvec$ direction inverts (\sctn{sctn:vortStrainrateMag}), it is seen that most-but-not-all $+1/2$ defects exist on $|\vortvec|=0$ (\fig{fig:Figure_1}f; dash-dotted lines). 
While this agrees with previous statements that defects tend to sit near the edges of vortices, our experiments and simulations demonstrate that this is not a strict colocalization constraint, since not every $+1/2$ defect sits on zero-vorticity contours (\fig{fig:Figure_1}f; top inset). 
Thus, it is not only where the vorticity is zero but rather where the vorticity and strain rate balance to produce simple shear. 

As a result, in both experimental and simulated results, $+1/2$ nematic defects reside directly on this flow boundary (\fig{fig:Figure_1}c). 
Likewise, $-1/2$ nematic defects are also most likely to be found on $\qcrit=0$; however, this is because they are initially created at $\qcrit=0$ due to the constraint on their $+1/2$ partner. 
In contrast to $+1/2$ defects, $-1/2$ defects are not constrained to remain on specific flow contours and can be associated with any value of $\qcrit$, existing in both strain-rate- and vorticity-dominated regions (\fig{fig:Figure_2}a). 
Thus, they have a wider range in their distance distribution, falling off exponentially with larger distances (\fig{fig:Figure_1}c). 
Indeed, the ideal picture of a solitary $-1/2$ defect is not located on $\qcrit=0$, but instead is encircled by a floweret of six vortices of alternating handedness (\fig{fig:Figure_2}b; \sctn{sctn:giomiPrediction}~\cite{giomi2014defect}). 
As a result, $-1/2$ defects reside on any sign of $\qcrit$-criterion (\fig{fig:Figure_2}a; \movie{mov:SimATCirculation}). 
In contrast, the striking coincidence of $+1/2$ defects on $\qcrit=0$ flow contours suggests an intrinsic relationship between flow structures and the trajectory of the self-propelled defects. 

\section{Spontaneous self-constraint violates the ideal, solitary view of defect-generated flow}

\begin{figure*}[t!]
    \centering
    \includegraphics[width=\columnwidth]{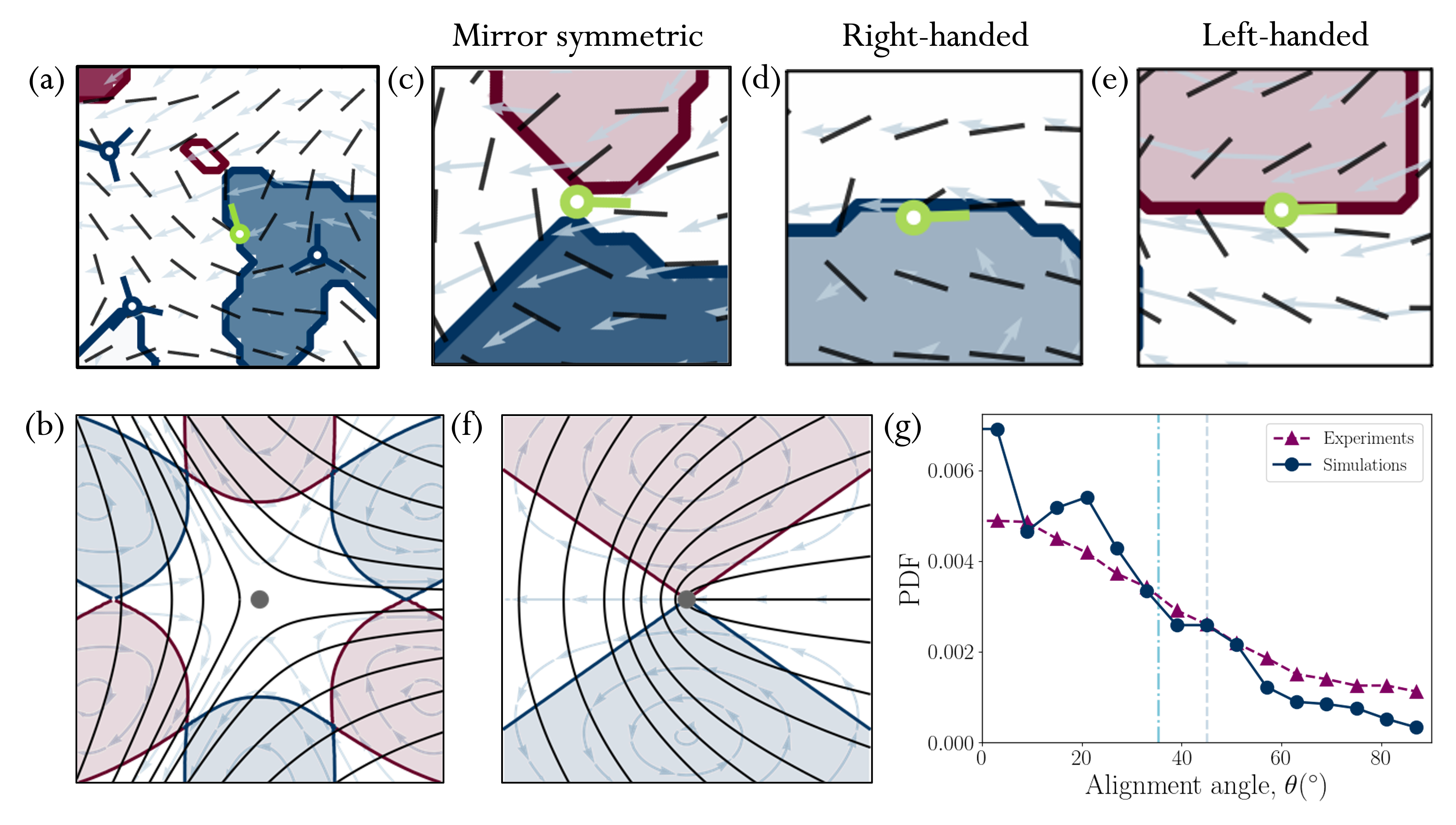}
    \caption{
    \textbf{Configurations of defects and viscometric surfaces.}
    \textbf{(a)} Snapshot of the flow geometry around minus and plus half nematic defects from the numerical simulations, illustrating three minus half defects in a strain-dominated region, in a vorticity-dominated region and on $\qcrit=0$ border. 
    Arrows show instantaneous velocity field, dashed lines director field, solid lines $\qcrit=0$ contours, with enclosed area colored by circulation (red for clockwise and blue for anti-clockwise). 
    Blue trefoil symbols mark $-1/2$ defects and  green comet-shaped symbols represent $+1/2$ defects. 
    \textbf{(b)} Prediction for ideal, solitary $-1/2$ defect (\sctn{sctn:giomiPrediction}). 
    \textbf{(c)-(e)} Snapshots of plus half defects, for the cases of
    \textbf{(c)} mirror symmetry, \textbf{(d)} mirror-symmetry-broken with anti-clockwise vortex, and \textbf{(e)} clockwise. 
    \textbf{(f)} Same as (b) for solitary $+1/2$ defect. 
    \textbf{(g)} Probability distribution function (PDF) of alignment angles between the defect orientation and the tangent of the associated viscometric ($\qcrit=0$) line (\sctn{sctn:interactions}). 
    The vertical blue dash-dot line indicates the ideally expected alignment from \textbf{(f)} (\sctn{sctn:giomiPrediction}) and the grey dashed line indicates the alignment angle for a point active force (\sctn{sctn:lineForceModel}).
    }
    \label{fig:Figure_2}%
\end{figure*}

Investigating the relationship between the viscometric ($\qcrit=0$) surfaces and the $+1/2$ defect positions reveals that the defect/surface complexes exist in two distinct configurations: 
Either a $+1/2$ defect is positioned at an intersection of viscometric lines (\fig{fig:Figure_2}c), or a defect lies parallel to a single $\qcrit=0$ viscometric line (\fig{fig:Figure_2}d-e). 
In the first case (\fig{fig:Figure_2}c), a $+1/2$ defect is positioned at a cross-roads between two viscometric lines, and has two equally strong counter-rotating vortex regions on either side of the defect axis, while the flow directly in front and behind is strain-rate dominated. 
Thus, the defect has a mirror symmetry along its head-tail axis. 
This corresponds to the ideal picture of a solitary defect (\fig{fig:Figure_2}f; \sctn{sctn:giomiPrediction}~\cite{giomi2014defect}), which is tacitly the expectation of flows around defects in active turbulence. 
The combination of these flow geometries ensures the velocity field flows parallel to the defect orientation and self-propulsion direction. 
However, this mirror-symmetric configuration is transient. 
Although the ideal model of solitary defects describes how active forces are a symmetric source of vorticity on both sides of the defects, in experiments and simulations actual vorticity-dominated regions are not self-propulsively moving along side self-motile defects as the picture might suggest. 

Instead, the mirror symmetry around the isolines is broken (\fig{fig:Figure_2}d-e). 
Each of the defects in these states orient parallel to a single viscometric line, with vorticity dominating on one side and strain rate dominating on the other. 
This spontaneous handedness of a subpopulation of defects is not predicted by the ideal model for an solitary defect; however, defects with this configuration persist for long durations and are observed to be the majority (\fig{fig:Figure_1}; \movie{mov:SimATCirculation})  because a motile defect can persistently move along the $\qcrit=0$ line, circulating around a single neighboring vortex. 
As a result, local chirality spontaneously emerges in this globally achiral system. 
This spontaneous handedness is necessarily erased when ensemble averaging fields around defect cores. 
The average flow profiles of the two non-symmetric cases are distinct from the mirror-symmetric case when the subpopulations are separately averaged (\figsi{fig:flowavrdefect}).

We quantify the distribution of these defect/viscometric surface complexes via the distribution between the defect alignment angles between the $+1/2$ defect axis and the $\qcrit=0$ boundary tangent (\fig{fig:Figure_2}g; \sctn{sctn:interactions}). 
In experiments and simulations, the most likely alignment is directly parallel, quantifying the preference for mirror-symmetry-broken configurations in active turbulence. 
This demonstrates that the mirror-symmetry-broken configuration is actually preferred to the symmetric configuration predicted by the ideal solitary-defect theory (\sctn{sctn:giomiPrediction}).
The simulations also show a secondary peak around $\theta\approx20\si{\degree}$, representing the minority-population in the transient symmetric configuration. 
The measured angle is smaller than $\theta\approx35.26\si{\degree}$, the ideal-model angle for a solitary defect (\sctn{sctn:giomiPrediction}). 
The experimental alignment angle distribution does not show a second peak, which may be due a combination of the $\qcrit$-criterion boundaries being less smooth or the mirror-symmetric case being even shorter-lived in experiments. 
Disclinations intermittently transform between the mirror-symmetric and the broken-mirror-symmetry cases (\movie{mov:SimATCirculation}). 
The results so far evidence that specific isolines of flow boundaries constrain the motion of topological defects but, in the next section, we show that there is a two-way interdependence and explain by what mechanism defects and viscometeric contours are pinned. 

\section{Bend walls and viscometric surfaces are interdependent}

\begin{figure*}[t!]
    \centering
    \includegraphics[width=1.0\columnwidth]{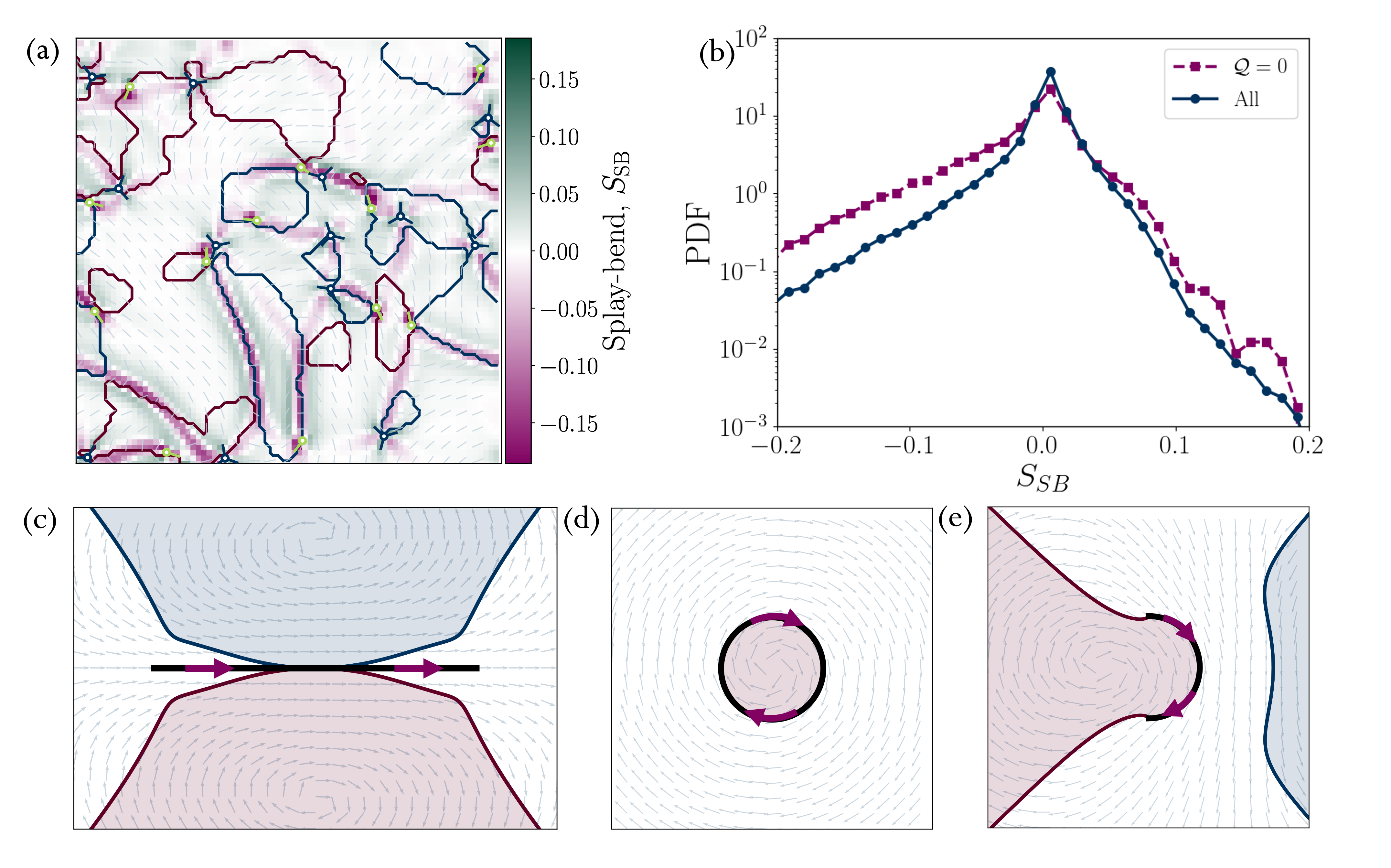}
    \caption{
    \textbf{Viscometric surfaces generated by bend walls.}
    \textbf{(a)} Same snapshot as \fig{fig:Figure_1}b with colormap corresponding to the scalar-bend bend parameter $\ssb$ (\sctn{sctn:ssb}). 
    Narrow lines of strong negative $\ssb$ are interpreted as bend walls. 
    \textbf{(b)} Probability distribution function (PDF) of $\ssb$ sampled equally for all points and times, against the distribution of $\ssb$ only on $\qcrit=0$ contours. 
    \textbf{(c)-(e)} $\qcrit$-criterion and velocity field predicted by Stokesian line force model of bend walls (\sctn{sctn:lineForceModel}).
    Red and blue shading indicate clockwise and anticlockwise vorticity-dominated regions, while the white regions are strain-rate dominated. The $\qcrit=0$ contours are solid lines.
    The unit-vector velocity field is shown with grey arrows. The line force is shown as thick solid black line with the direction of forcing indicated by the purple arrows. Line force taken as:  
    \textbf{(c)} finite line, 
    \textbf{(d)} full-circle and 
    \textbf{(e)} half-circle. 
    }
    \label{fig:Figure_3}%
\end{figure*}

Though topological defects are often identified as dominant localized-sources of active forcing (\sctn{sctn:giomiPrediction}), more generally force is generated from any deformation of the orientation field (\sctn{sctn:sims}). 
In particular, \textit{bend walls} play an essential role in extensile active nematics~\cite{martinez2019}. 
Bend walls are narrow lines of high bend deformation forming sharp kinked lines between nematic domains, representing a nematic N\'{e}el wall~\cite{Patelli2019}. 
As highly localized deformations, bend walls generate strong active forcing and, because extensile nematics are hydrodynamically unstable~\cite{Aditi2002}, bend constricts into system-spanning narrow kink lines before defect pairs unbind along these walls and the self-motile $+1/2$ defects unzip the bend wall as it advances~\cite{thampi2014instabilities}. 
After this initial development of active turbulence, it is then presumed that significant forces only occur in the immediate vicinity of defects. 
However, in practice this is not observed in experiments or simulations. 
Rather, the hydrodynamic instability is incessantly constricting bend into sharp kink walls, as demonstrated by the splay-bend parameter $\ssb$, which indicates the difference between splay and bend deformations (\sctn{sctn:ssb}). 
The splay-bend parameter traces bend walls through strongly negative $\ssb$ values (\fig{fig:Figure_3}a; \movie{mov:SimATssb}), and reveals that bend walls correlate strongly with the $\qcrit=0$ viscometric lines. 
The distribution of $\ssb$ coinciding with viscometric surfaces is more skewed towards negative (bend) values than for the system as a whole (\fig{fig:Figure_3}b).

\subsection{Physical Model}
The extended concurrence of bend walls and $\qcrit=0$ lines suggests the physical mechanism by which defect dynamics and viscometric surfaces become cross-constrained. 
We posit that the elongated bend walls are crucial to the pure shearing flows of the viscometric surfaces.
To understand how these mesoscale nematic and hydrodynamic structures are intertwined, we consider a series of simple models. 
These models will demonstrate that the bend walls linking $\pm1/2$ defects (\fig{fig:Figure_3}a) explain both the viscometric contour and the handedness of elastically bound $+1/2$ defects, as long as the walls are \textit{(i)} finite in length and \textit{(ii)} follow curved paths. 

We have already seen how the ideal analytical solution of a solitary $+1/2$ defect explains the mirror-symmetric case but not the mirror-symmetry-broken case (\fig{fig:Figure_2}). 
In fact, even modelling solitary defects as simple point-forces is sufficient to qualitatively predict mirror symmetric defects at $\qcrit=0$ intersections (\sctn{sctn:lineForceModel}). 
A straightforward model that accounts for the elongated nature of bend walls might treat them as infinitely long, perfectly straight structures (\sctn{sctn:straightwall}). 
Indeed, this is consistent with $\qcrit=0$; however, this overly simplified model predicts $\qcrit=0$ everywhere, rather than just on the bend wall. 
This is because perfectly straight, infinitely long bend walls produce only shear. 

To address this shortcoming, bend wall curvature is added to the model by perturbing the bend walls into infinitely long sinusoidal undulations. 
This model numerically demonstrates that curved bend walls produce closed viscometric surfaces (\figsi{fig:wavybendwall}).
Furthermore, this model demonstrates that active viscometric flows drive additional bend constriction, narrowing the bend walls into kink lines but not further perturbing their conformation (\sctn{sctn:wavywalls}). 
This re-emphasizes an essential property: 
$\qcrit=0$ lines, where vorticity and strain-rate balance, is also where velocity gradients exhibit no stretching or elongation of equidistant streamlines. 
This is not only why $\qcrit=0$ is referred to as viscometric, but also why the bend walls are not further perturbed by the flows and why flows do not misalign defects from following $\qcrit=0$ lines. 
Thus, this simple model suggests that for the bend walls to be deformed, they cannot be infinitely long. 

These results justifies an even simpler approach of modeling $+1/2$ defects unzipping narrow finite-length kink walls as idealized lines of active force. 
We assume active stresses dominate elastic stresses, and hence, model the flows using Stokes equation (\sctn{sctn:lineForceModel}). 
The bend direction, and therefore active force, is parallel with the tangent of the bend wall. 
By solving the Stokes equation, we calculate the velocity field and $\qcrit=0$ contours for the following cases: \textit{(i)} a finite straight line,  \textit{(ii)} a perfect circle, and \textit{(iii)} an arc (\fig{fig:Figure_3}c-e). 
Case \textit{(i)} assumes a defect is the end of a finite straight kink wall, which accentuates the acute angle (\fig{fig:Figure_3}c; \sctn{sctn:lineForceModel}). 
This explains why the observed alignment angle (\fig{fig:Figure_2}g) is substantially smaller than the ideal solitary value. 
However, this bend wall conformation retains the mirror symmetry of the flow field --- in the limit of an infinitely long line, the two $\qcrit=0$ lines collapse into a single line but simultaneously $\qcrit$ goes to zero everywhere. 
Case \textit{(ii)} instead considers a perfectly circular bend wall (\fig{fig:Figure_3}d) for which a $\qcrit=0$ contour encircles a vorticity-dominated core, with a strain-rate dominated exterior. 
While completely circular bend walls are not observed in experiments or simulations, this model produces a well-defined vortex, which is the basis of the current understanding of energy spectra in active turbulence~\cite{giomi2015}. 
The intermediate case \textit{(iii)} of arced bend walls explains the symmetry-broken regime in which many of the defects exist. 
Once defects are present in the system, bend walls continuously curve (\fig{fig:Figure_3}a). 
To simply model curved bend walls, we numerically solve an arced force line (\fig{fig:Figure_3}e). 
This model reproduces two viscometric contours, but they no longer intersect. 
Rather the viscometric line on the concave side of the arc continues to coincide with the bend wall, while the $\qcrit=0$ contour on the convex side separates, which is the physical picture observed in active turbulence (\fig{fig:Figure_1}).

Hence, the straightforward model of a finite, curved force line explains how bend walls play the crucial role in setting the interdependence of defect dynamics and viscometric surfaces: 
First, bend constriction via the hydrodynamic instability causes active forces to be localized along narrow lines, resulting in shearing flows that neither stretch nor elongate streamlines. 
Self-motile $+1/2$ defects unzip these narrow bend walls and thus are elastically constrained to $\qcrit=0$. 
Finally, the finite length and curvature of the bend walls causes the mirror-symmetry-broken case. 

\section{Beyond extensile, flow-aligning 2D active nematic turbulence}

\begin{figure*}[t!]
    \centering
    \includegraphics[width=\columnwidth]{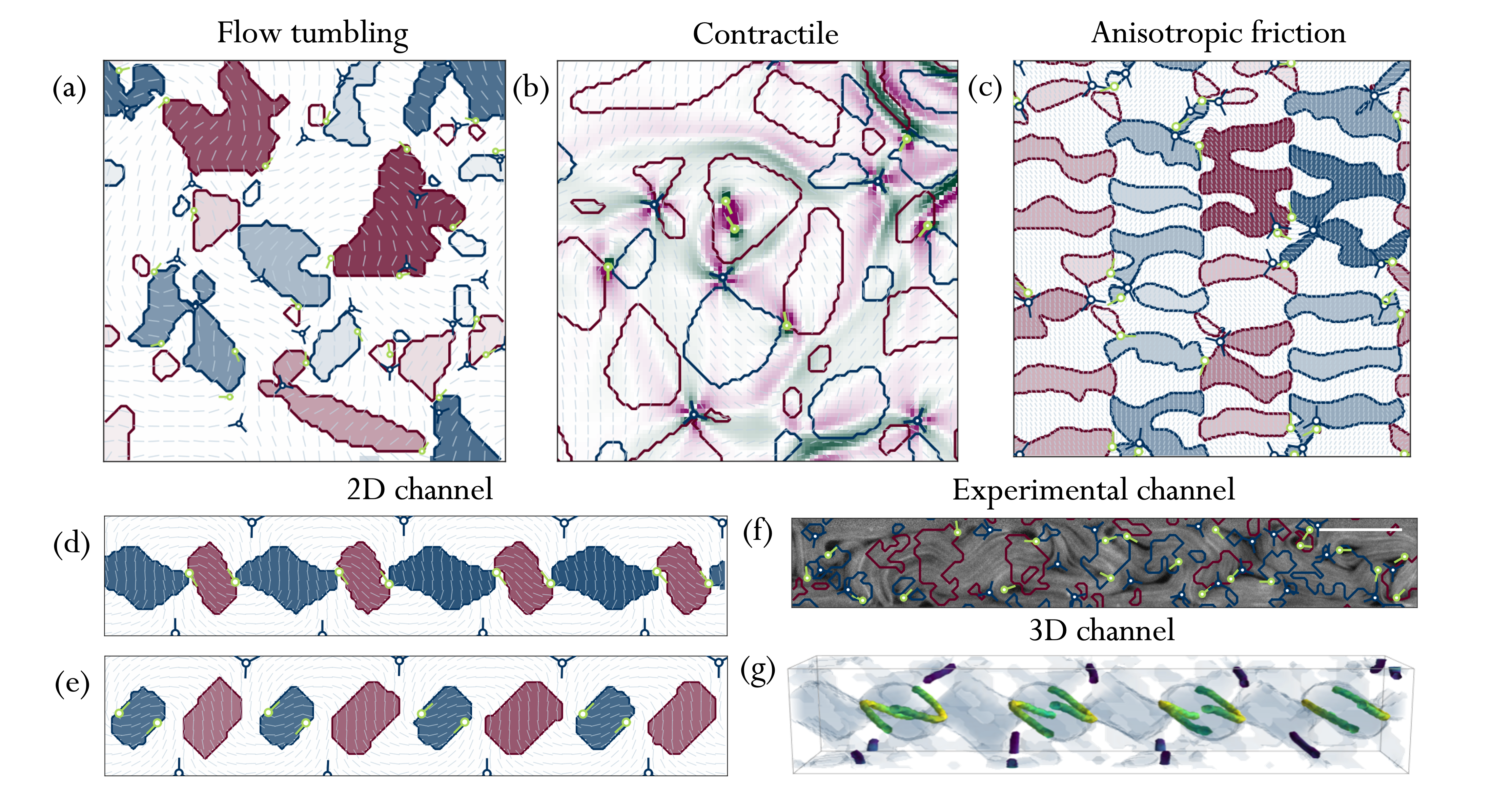}
    \caption{
    \textbf{The spontaneous self-constraint between motile $+1/2$ defects and viscometric surfaces holds more generally.}
    \textbf{(a)} Active extensile turbulence with an alignment parameter in the flow-tumbling regime. 
    \textbf{(b)} Contractile active turbulence, in an operating regime with rare instances of bound $+1/2$ defect pairs creating effective $+1$ defects, which do not need to be on $\qcrit=0$. 
    Same visualization as \fig{fig:Figure_3}a.
    \textbf{(c)} Lanes of alternating circulation form with anisotropic friction in the flow-aligning regime. 
    \textbf{(d)-(e)} Dancing defect dynamics in 2D confining channel, with $+1/2$ defects switching between vortices and at $\qcrit=0$ intersections in (d) and aligned along vortex boundaries in (e). 
    (a)-(e) show numerical simulations of 2D active nematics, with colormaps, lines and markers the same as \fig{fig:Figure_1}b. 
    \textbf{(f)} Experimental verification that the spontaneous self-constraint is retained in a channel confinement. Scale bar: \SI{100}{\micro\meter}.
    \textbf{(g)} Three-dimensional vortex lattice with disclinatines colored by the twist angle ($\cos\beta$ from \sctn{sctn:defectMethods}). 
    Yellow indicates local $+1/2$ wedge profile, green twist-type profile and blue $-1/2$ wedge profile. 
    Grey shading illustrates vorticity-dominate regions ($\qcrit>0$), and white where strain-rate dominates ($\qcrit<0$). 
    }
    \label{fig:Figure_4}%
\end{figure*}

This study has focused on the spontaneous self-constraint between defects and flow structures in unconfined 2D extensile nematic turbulence. 
However, our evidence suggests this phenomenon is substantially more general, apparently holding whenever $+1/2$ topological defects are present in active nematics (\fig{fig:Figure_4}). 

The conclusions arrived at for flow-aligning active turbulence also hold in the \textit{flow-tumbling} regime (\fig{fig:Figure_4}a). 
Once again, defects reside on viscometric contours, and either travel along a single vortex boundary or transiently cross an intersection (\movie{mov:SimFT}). 
The interdependence strongly resembles the flow-aligning regime, which is consistent with the expectation that active turbulence in flow-aligning and flow-tumbling behaves similarly. 
Futhermore, the conclusions likewise generally hold for \textit{contractile} activity (\fig{fig:Figure_4}b). 
Minus half defects are still non-motile and unbound from viscometric lines. 
Motile $+1/2$ defects now chase their tail (move in the opposite direction compared to extensile defects) and unzip \textit{splay walls}; yet, they are still most likely to be found self-propulsively moving where $\qcrit=0$ (\movie{mov:SimContractile}). 
However in flow-aligning contractile systems, there is an intriguing exception: 
In a narrow parameter regime, a pair of similarly charged $+1/2$ defects can temporarily form impermanent $+1$ complexes~\cite{thijssen2020binding}. 
These effectively immotile $+1$ complexes produce a flow profile that does not necessitate $\qcrit=0$ and, thus, these effectively $+1$ compounds do not reside where $\qcrit=0$, allowing them to briefly elude the confinement condition. 
Thus, self-motility of $+1/2$ defects, actively unzipping deformation walls, is central to the reported interdependence. 

How robust is the interdependence between $+1/2$ defects and $\qcrit=0$ to extrinsic conditions? 
Homogeneous friction, for example, can modify the properties of active turbulence~\cite{thijssen2021submersed,MartinezPrat2021} and anisotropic friction can order flow states~\cite{guillamat2017,thijssen2020active}. 
\textit{Anisotropic friction} creates an easy flow axis, forming lanes with preferred circulation handedness encircled by $\qcrit=0$ (\fig{fig:Figure_4}c; \movie{mov:SimAniFric}). 
The vortex-dominated regions form a lattice of viscometric contours, with intersection points along the boundary at advective lanes.
Defects still reside on $\qcrit=0$ and, moreover, now appear to be predominantly found on intersections. 
This suggests a distinction between the low and high friction regimes; defects are more likely to be in the mirror-symmetry-broken configuration without friction, while these high friction regimes combat bending curvature/instabilities, which our theoretical model suggests will make mirror-symmetric defects more likely. 

Impermeable, no-slip boundary conditions can also have profound effects on spontaneous active flow states, creating complex spatio-temporally ordered dynamics, even in simple geometries, such as 2D \textit{channel confinement}~\cite{shendruk2017dancing,hardouin2019}. 
The resulting vortex lattice and associated defect dancing is clear in simulations (\fig{fig:Figure_4}(d,e); \movie{mov:SimVLCirc}), but also apparent in experiments (\fig{fig:Figure_4}f; \movie{mov:ExpVL}). 
In both simulations and experiments, the $+1/2$ defects orient parallel to the viscometric line, when traversing a vortex (\fig{fig:Figure_4}e), and are able to cross to neighboring vortices by passing through the fleeting $\qcrit=0$ intersection (\fig{fig:Figure_4}d; \movie{mov:ssbVL}). 
Similarly, in experiments $+1/2$ defects weave through the center, never departing from $\qcrit=0$ and transversing to neighboring vortices through viscometric intersections. 
These dancing dynamics beautifully illustrate the transitions between the mirror-symmetric and mirror-symmetry-broken configurations (\figsi{fig:ceilidh}). 
Plus-half defects orient and travel parallel to the edge of a vortex, until the vortices instantaneously touch forming a short-lived $\qcrit=0$ intersection, allowing the defects to transfer to the next vortex of opposite handedness---all the while, the defects remain on the $\qcrit=0$ lines. 

In fact, the interdependent spontaneous self-constraint holds in \textit{three-dimensions}, where the point defects become disclination lines and viscometric contours become 2D surfaces. 
The vortex lattice exists in a 3D duct~\cite{keogh2022helical}. 
Analogously to the two-dimensional case, wedge-type $-1/2$-profile disclination lines exist at the walls, while disclinations with a wedge-type $+1/2$ profile dance around the central vortex lattice (\fig{fig:Figure_4}g). 
The disclinations at the center with wedge-type $+1/2$ profiles continually maintain contact with the two-dimensional $\qcrit=0$ isosurface (\movie{mov:3DVL}), demonstrating that motile defects are constrained to lie on $\qcrit=0$ surfaces even in 3D.
Altogether, the examples explored in \fig{fig:Figure_4} indicate that the self-emergent constraint between motile defect dynamics and viscometric surfaces is not only a property of extensile active turbulence but is more general, apparently holding whenever motile defects are present in active nematics. 

\section{Conclusion}\label{sctn:conc}

In conclusion, we have identified and explained a spontaneously arising constraint between motile defect dynamics and viscometric surfaces, where vorticity and strain-rate balance. 
This work challenges the idea that nematic defects are solely responsible for their own dynamics --- ultimately, neither defects nor hydrodynamics alone govern the multi-field dynamics that spontaneously arise in active nematics; rather, they are intrinsically interdependent. 
Identifying this spontaneous self-constraint revealed that the ideal picture of solitary self-motile defects generating a pair of mirror-symmetric vortices is not strictly true in practice. 
Instead, motile defects more often move along a single viscometric surface, where there is no stretching nor elongation of the fluid, a fact that appears to hold for all motile defects in active nematics and not only 2D extensile active turbulence. 
This shows that defects can be classified into three conformations based on local handedness with respect to their viscometric surface surroundings. 
Until now, this variation, as well as any possible different properties, had been negelected. 
Not only do the hydrodynamics not force the plus-half defects off these contours, but they coincide with the very bend walls that these defects are unzipping --- the field dynamics are codependent. 
Furthermore, these results underscore the continual role of bend wall constriction and unzipping in steady state dynamics, and highlight the centrality of mesoscale structure in collective dynamics, providing a mechanism to manipulate the degree of orderly dynamics in a system. 

We anticipate that our observation of co-dependence will aid efforts to understand the highly complicated structure of 3D active turbulence~\cite{kralj2023}, and complement recently developed Lagrangian descriptions of coherent structures~\cite{serra2023}. 
The self-constraint identified here could lead to designs for novel active functionality~\cite{zhang2021,zhang2022} or biointerfaces~\cite{Zhang2022b}. 
Biological systems, including active transport in microbial colonies~\cite{dhar2022}, mitotic spindle assembly~\cite{needleman2017} tissue responses~\cite{xi2019}, epithelial reorganisation~\cite{henkes2020} and morphogenesis~\cite{Vafa2022}, all involve a careful balance between collective motion and protected-structures for their functionality. 
Ultimately, spontaneous constraints in active materials could have far-reaching implications as a framework for understanding how biological systems employ active stresses for simultaneous dynamics and restraint.


\clearpage
\section{Method section}\label{sctn:methods}

\subsection{Experiments}\label{sctn:exp}

\subsubsection{Materials}
Short and stable microtubules (MTs) were polymerized by incubating, at \SI{37}{\celsius} for 30 minutes, a mixture containing \SI{8}{\milli\gram\per\milli \liter} of recycled tubulin from bovine brain (Brandeis Materials Research Science and Engineering Center), \SI{6}{\milli \molar} of guanosine-5'-[($\alpha$,$\beta$)-methyleno]triphosphate (GMPCCP, Jena Biosciences, NU-405S), and \SI{1}{\milli \molar} of DL-dithiothreitol (DTT, Sigma, 43815) in M2B buffer (\SI{80}{\milli \molar} of PIPES [Sigma, P1851], \SI{1}{\milli \molar} of EGTA [Sigma, E3889], \SI{2}{\milli \molar} of \ce{MgCl2}, pH=6.8). After incubation, the solution was kept at room temperature for 5 hours, then frozen in liquid nitrogen and stored at \SI{-80}{\celsius} for future use. For confocal fluorescence microscopy, 3\% of the tubulin was labelled with Alexa-647, a bright, far-red–fluorescent dye.
Biotinylated kinesin was prepared in the BioNMR group at the institute of bioengineering of Catalunya (IBEC). Drosophila melanogaster heavy-chain kinesin-1 K401-BCCP-6His was expressed in \textit{Escherichia coli} using the plasmid WC2 from the Gelles Laboratory (Brandeis University). Biotinylated kinesins were purified with a nickel column and dialysis against \SI{500}{\milli \molar} imidazole aqueous buffer. The kinesin concentration was estimated at \SI{2.5}{\micro\molar} by absorption spectroscopy, and was finally stored in a 40\%w/v sucrose solution at \SI{-80}{\celsius} for future use. Kinesin clusters were obtained by preparing a mixture with \SI{1}{\micro\molar} of biotinylated kinesin dimers and \SI{31}{\micro\gram\per\milli\liter} of tetrameric streptavidin in a M2B buffer supplemented with \SI{0.2}{\milli\molar} of DTT, this stoichiometric ratio corresponding approximately to 2 kinesin dimers per cluster. This mixture was incubated on ice for 30 minutes.

The active microtubule/kinesin gel consisted in a M2B preparation with \SI{1.6}{\milli\gram\per\milli\liter} of MTs, motor clusters (the concentration of streptavidin was \SI{8.2}{\micro\gram\per\milli\liter}), \SI{1.4}{\milli\molar} of adenosine triphosphate (ATP, Sigma, A2383) to fuel the molecular motors and 1.6\% w/v of depleting agent polyethylene glycol (PEG, 20 kDa, Sigma, 9517) to induce the bundling of the MTs. An ATP-regenerating system (2.8\% v/v of pyruvate kinase/lactate dehydrogenase [PK/LDH, Sigma, P02] and \SI{26.2}{\milli\molar} of phosphoenolpyruvate [PEP; Sigma; P7127]) was employed to keep a fixed concentration of ATP during the experiments. To prevent photobleaching and oxidation, some antioxydants were included in the active gel: \SI{5.4}{\milli\molar} of DTT, \SI{3.3}{\milli\gram\per\milli\liter} of glucose (Sigma; G8270), \SI{38}{\micro\gram\per\milli\liter} of catalase (Sigma, C40), \SI{0.22}{\milli\gram\per\milli\liter} of glucose oxidase (Sigma G2133) and \SI{2.0}{\milli\molar} of trolox (Sigma, 238813). The \ce{MgCl2} concentration was raised with 4.7\% v/v of Mix solution (\SI{69}{\milli\molar} of \ce{MgCl2} in M2B). The pH of the active gel was adjusted to 6.8. 

\subsubsection{Active nematic assembly}

Active nematic layers were assembled at a water/oil interface following two different protocols. The unconfined active nematic layer (as shown in \fig{fig:Figure_1}a) was prepared in a closed observation chamber, consisting of a bottom hydrophobic glass slide treated with Aquapel and a hydrophilic cover slip coated with a polyacrylamide brush, separated by \SI{120}{\micro\meter} of double-sided tape, forming a rectangular chamber of size \SI{3}{\milli\meter}$\times$\SI{22}{\milli\meter}. The chamber was first filled with fluorinated oil HFE7500 with 1.8\% v/v of 008-FluoroSurfactant (RAN Biotechnologies), then \SI{10}{\micro\liter} of the active gel was introduced by capillarity in the chamber, replacing most of the oil but letting a lubricating film of oil at the surface of the bottom glass slide. The chamber was sealed with UV-curable glue (Norland NOA-81). The assembly of the microtubule bundles at the water/oil interface was subsequently sped up by centrifuging the sample 10 minutes at 215$\times$g. For laterally confined active nematics (as shown in \fig{fig:Figure_4}f), we prepared an open observation chamber by sticking with UV-curable glue circular polydimethylsiloxane (PDMS) walls (\SI{1}{\centi\meter} in diameter) to a polyacrylamide coated glass slide. \SI{2.5}{\micro\liter} of the active gel supplemented with 2.2\% w/v of surfactant pluronic F-127 (Sigma, P-2443), was deposited in the well and spread on the hydrophilic glass surface, then was immediately covered with \SI{200}{\micro\liter} of silicone oil with a dynamic viscosity of \SI{20}{\milli\pascal\second}. The assembly of the microtubules bundles at the water/oil interface occurred on their own within 40 minutes. 

\subsubsection{Micro-printed grid walls}\label{sctn:expChannel}
To confine laterally the active nematic to channels, as shown in \fig{fig:Figure_4}f, we used millimeter-sized platforms in photoresist encompassing the rectangular enclosures. Such grids were introduced in the open observation chamber from the top and placed at the water/oil interface, consequently trapping the AN in the rectangular channels~\cite{hardouin2019}. Grids were 3D printed with a two-photon polymerization printer, a Nanoscribe GT Photonic Professional device, with a negative-tone photoresist IP-S (Nanoscribe GmbH, Germany) and a $25\times$ objective. The grids were directly printed on silicon substrates without any preparation to avoid adhesion of the resist to the substrate. After developing 30 minutes in Propylene Glycol Monomethyl Ether Acetate (PGMEA 99.5\%, Sigma, 484431) and 5 minutes in isopropanol, a batch polymerization was performed with UV exposure (5 minutes at 80\% of light power). The pattern resolution achieved was about \SI{50}{\nano \meter}. The grid thickness was \SI{100}{\micro\meter}, to ensure a good resistance during manipulation.

\subsubsection{Imaging and image processing}\label{sctn:imaging}
Images of the active nematic layer were acquired with a Nikon eclipse Ti-E inverted microscope, equipped with a confocal spinning disk CSU-X1 (Yokogawa) and a motorized stage Nikon Ti-S-EJOY, and a $10\times$ objective (Nikon). The sample was excited with a laser beam at \SI{647}{\nano \meter} with an exposure between 200-500 \si{\milli \second} depending on the sample fluorescence. Images were captured with a camera Hamamatsu Orca Flash4.0, and acquired with the software NiS-Elements. The acquisition frame rate was 2 frame per second.

Image processing, including background correction, was carried out with the software ImageJ. 
The nematic tensor $\nemT\spacetime$ was computed from the images using custom Matlab scripts~\cite{ellis2018} 
that infer local alignment of the microtubules using coherence-enhanced diffusion filtering. First, noise in the image is filtered out with a Gaussian blur of standard deviation $\sigma_1$. The pixel-level orientation is obtained by finding the direction along which the fluorescence intensity is most homogeneous and smoothed by means of a Gaussian blur of standard deviation $\sigma_2$. Finally, the nematic tensor $\nemT$ is constructed by ensemble averaging the molecular orientation tensor over a small circle of size $\beta$ around each pixel. The three parameters $\sigma_1$, $\sigma_2$ and $\beta$ were manually adjusted by visual inspection of the resulting director field. Velocity fields $\velV\spacetime$ were calculated with the optical flow method detailed in ref.\cite{SunSOOF} and associated Matlab codes, using the 'classic+nl-fastp' method. 

The high resolution images exhibit pixel-level substantial noise, which hinders the calculation of gradients of the fields, required for the identification of defects and viscometric surfaces. 
Therefore, we generate a coarse-grained field which only takes the value once every chosen number of pixels. 
The same resolution was also used for the velocity fields for the velocity-gradient based analysis. 
We test the ideal resolution using the defect tracker (\sctn{sctn:anal}) and choose a value that identifies the most number of defects correctly (while avoiding overestimating). 
For both bulk active turbulence and channel confinement, the chosen resolution was 16. 

\subsection{Simulations}\label{sctn:sims}
Our active nematic simulations employ a hybrid lattice Boltzmann approach~\cite{orlandini2008hydrodynamic} to reproduce the continuum nematohydrodynamics model~\cite{thampi2016active}.
The dynamics of the orientational order parameter $\nemT\spacetime=S\left(\dirV\otimes\dirV-\unitT/3\right)$ at each position $\posV$ and time $\time$ in 3D is described by the Beris-Edwards transport equation
\begin{align}
    \label{eq:nem}
    \mathcal{D}_t \nemT &= \Gamma \molFieldT, 
\end{align}
where the right-hand side is the relaxation to equilibrium through a relaxation parameter $\Gamma$ and molecular field $\molFieldT = -\left( \frac{\delta F}{\delta\nemT} - \frac{1}{d}\unitT \; \tr{ \frac{\delta F}{\delta\nemT} } \right)$ is the molecular field corresponding to the nematic free energy $F$. 
On the left-hand side, $\mathcal{D}_t \nemT = \left( D_t  - \coRot \right)\nemT$ is a co-variant derivative that accounts for material advection by $D_t=\partial_t + \velV\cdot\grad$, which couples directly to the velocity field, and a corotational operator $\coRot\left\{\nemT\right\} = \left\{ \lambda \strainrateT+\vortT,\tilde{\nemT} \right\}_+ - 2\lambda\left(\nemT:\gradVelT\right)\tilde{\nemT}$. 
The co-rotational term couples the orientation to the velocity gradient $\gradVelT=\grad\otimes\velV$ through the vorticity $\vortT=\left(\gradVelT-\transpose{\gradVelT}\right)/2$ and strain-rate $\strainrateT=\left(\gradVelT+\transpose{\gradVelT}\right)/2$ tensors, written using a (anti)commutator $\left\{\tens{A},\tens{B}\right\}_\pm = \tens{A}\cdot\tens{B} \pm \tens{B}\cdot\tens{A}$, with a flow-aligning parameter $\lambda$ and $\tilde{\nemT}\equiv\nemT+\unitT/d$. 
The Beris-Edwards equation (\eq{eq:nem}) is numerically solved by finite difference. 

The velocity $\velV\spacetime$ evolves according to the generalized Navier-Stokes equations
\begin{align}
    \label{eq:navier_stokes}
    \dens D_t \velV &= \div \stressT - \frictionT \cdot \velV
\end{align}
with density $\dens$ and friction tensor $\frictionT$. 
The stress $\stressT$ consists of a viscous term $2\visc \strainrateT$ with viscosity $\visc$, a pressure term $-p \unitT$, an elastic term $K\left[ 2\lambda\left(\nemT:\molFieldT\right)\tilde{\nemT} - \lambda \left\{\molFieldT,\tilde{\nemT}\right\}_{+} + \left\{\nemT,\molFieldT\right\}_{-} - \grad \left( \nemT: \tfrac{\delta F}{\delta\grad\nemT} \right) \right]$, and, most relevantly an active term, 
\begin{align}
    \label{eq:activeStress}
    \stressT_\act &= -\act \nemT.
\end{align} 
Both the elastic and active terms couple to velocity field to the nematic field, but the active stress is directly proportional to the nematic field, with an activity coefficient $\act$ that leads to a bend instability when $\act>0$ (or a splay instability when $\act<0$), which drives deformation, wall formation and defect unbinding, leading to the active defect turbulence. 
The lattice Boltzmann algorithm simulates \eq{eq:navier_stokes} and achieves near incompressiblity $\div\velV \approx 0$. 

The free energy includes a bulk component and a surface anchoring term $F=\int_V f \ dV +\int_A f_A \ dA $
The bulk free energy density is constructed as a Landau-de Gennes expansion with an one-constant elastic Oseen-Frank term 
\begin{align}
    f &= A_0\left[ \left(\frac{1-\gamma/3}{2}\right) \nemT:\nemT -\frac{\gamma}{3} \nemT:\left(\nemT\cdot\nemT\right) +\frac{\gamma}{4}\left(\nemT:\nemT\right)^2 \right] + \frac{K}{2}\left(\grad\otimes\nemT\right)	\vdots \left(\grad\otimes\nemT\right)
\end{align} 
with $\gamma$ controlling the distance from the isotropic-nematic phase transition and $K$ the isotropic elastic constant. 
The surface free energy density is 
\begin{align}
    f_A &= W_1\left(\nemT-\nemT^\perp\right):\left(\nemT-\nemT^\perp\right) + W_2 \left(\nemT:\nemT - S_\text{eq}^2 \right), 
\end{align} 
where $W_{1,2}$ are anchoring energies, $6S_\text{eq}=1+3\sqrt{1-8/3\gamma}$ is the equilibrium scalar order parameter and $\nemT^\perp = \tens{P}\cdot\left(\nemT+S_\text{eq}\unitT/3\right)\cdot\tens{P}$ for the surface normal projection operator $\tens{P}$. 

Except where otherwise stated, our 2D extensile nematic simulations use the parameters (in lattice Boltzmann units) density $\dens=1$, rotational diffusivity $\Gamma = 0.3375$, flow-aligning parameter $\lambda = 1$, bulk energy density scale $A_0=1$, distance from the isotropic-nematic transition $\gamma=3$, dynamic viscosity $\visc=4/3$, Frank elasticity $K=0.05$, friction $\frictionT=\tens{0}$ and extensile activity $\act=0.1$. 
Periodic boundary conditions are used in a system of $80\times80$. The velocity and director fields are randomly initialized. 
A warmup time of $T_W=1\times10^4$ is used before data is collected for $T=2\times10^4$ lattice Boltzmann times steps. 

In \fig{fig:Figure_4} we show that the conclusions arrived at for 2D extensile active turbulence hold under other conditions. 
As far as possible, we attempt to keep our simulation parameters unchanged from the extensile case and only make a few substitutions. 
\begin{description}
    \vspace{-0.5\baselineskip}
    \item[Flow tumbling (\fig{fig:Figure_4}a; \movie{mov:SimFT})] Substitute $\lambda=0.6$.
    \vspace{-0.5\baselineskip}
    \item[Contractile (\fig{fig:Figure_4}b; \movie{mov:SimContractile})] Substitute $\act=-0.1$ and system size $160\times160$. We note that these contractile simulations permit a small non-zero component to the out-of-plane director and velocity field. This escape into the third dimension is negligible everywhere except for a small region at the top of \movie{mov:SimContractile}, which persisted in time but did not grow.
    \vspace{-0.5\baselineskip}
    \item[Anisotropic friction (\fig{fig:Figure_4}c; \movie{mov:SimAniFric})] We substitute $\frictionT = \left[ \left[ 0.015, 0\right],\left[ 0, 0 \right] \right]$. 
    \vspace{-0.5\baselineskip}
    \item[2D channel confinement (\fig{fig:Figure_4}d-e; \movie{mov:SimVLCirc})] Periodic boundary conditions kept at the ends of the channels but the channel walls are impermeable, no-slip boundaries with anchoring energy $W_1=W_2=0.3$. 
    System dimensions $130\times25$. 
    \vspace{-0.5\baselineskip}
    \item[3D channel confinement (\fig{fig:Figure_4}g; \movie{mov:3DVL})] Same as the 2D channel wall but now in a 3D square duct. 
    \item[\figsi{fig:flowavrdefect}] We substitute density $\dens=40$, rotational diffusivity $\Gamma=0.05$, dynamic viscosity $\visc = 1/6$, activity $\act=0.05$, system size $256 \times 256$, warmup time $T_W=5\times10^2$ and data collection time $T=1.95\times10^4$ lattice Boltzmann time steps. 
\end{description}

\subsection{Analysis}\label{sctn:anal}

\subsubsection{Defect Analysis}\label{sctn:defectMethods}
Topological disclinations in both experiments and 2D simulations are identified by calculating the topological charge density~\cite{schimming2022,Blow2014,DellArciprete2018} 
\begin{align}
    \label{eq:charge}
    q &= \frac{1}{4\pi}(\partial_x Q_{x\alpha} \partial_y Q_{y\alpha}-\partial_xQ_{y\alpha}\partial_yQ_{x\alpha}), 
\end{align}
using summation convention on Greek indices. 
Points where $q$ is greater than (less than) $1.5$ ($-1.5$) standard deviations are identified as $k=+1/2$ ($-1/2$) defects. 
Once each 2D defect is identified, its orientation is found via the angle 
\begin{align}
    \label{eq:defectAngle}
    \psi &= \frac{k}{k-1}\arctan\left(\frac{\langle \sgn{k} \partial_x Q_{xy}-\partial_yQ_{xx}\rangle_\circ}{\langle \partial_x Q_{xx} + \sgn{k}\partial_yQ_{xy}\rangle_\circ} \right),
\end{align}
with $\langle \cdot \rangle_\circ$ denoting averaging a closed loop around the disclination~\cite{vromans2016orientational}. 
In 3D, defect lines are identified and characterised using the disclination density tensor~\cite{schimming2022}, $\disclinDensT$, constructed from gradients in the nematic $\nemT$-tensor. 
In index notation, $\disclinDens_{ij}=\epsilon_{i\mu\nu}\epsilon_{jlk}\partial_l\nemOrder_{\mu\alpha}\partial_k\nemOrder_{\nu\alpha}$. 
The tensor is interpreted as $\disclinDensT=s\vec{\mathcal{R}}\otimes \vec{T}$, in terms of a non-negative scalar field $s$, the rotation vector $\vec{\mathcal{R}}$, and tangent vector $\vec{T}$ along the disclination. 
The values of $s$ approach a maximum at defect cores. We identified the defect lines as isosurfaces where $s=0.1$. 
We characterise the local geometry of the disclination line through the angle between the rotation and tangent vectors $\cos\beta = \vec{\mathcal{R}}\cdot\vec{T}$, where $\beta$ is known as the twist angle. 
Values of $\cos\beta$ smoothly transition between $-1$ for minus-half wedge profiles, through to $+1$ for plus-half wedge profiles. 
Visualisations in 3D used the Mayavi python library~\cite{ramachandran2011}.

\subsubsection{$\qcrit$-criterion}\label{sctn:QCrit}
We characterize the principle behavior of flow fields by the $\qcrit$-criterion, which is defined to be the second invariant of the velocity gradient $\gradVelT=\grad\otimes\velV$~\cite{jeong1995}
\begin{align}
    \qcrit &= \frac{1}{2}\left( \left(\tr{\gradVelT}\right)^2 - \tr{\gradVelT^2} \right) . 
    \label{eq:QcritSecond}
\end{align}
In the incompressibile limit, $\tr{\gradVelT}=\div\velV = 0$ and so this definition reduces to \eq{eq:Qcriterion}), which we repeat here for convenience
\begin{align}
    \qcrit &= \frac{1}{2}\left( \lVert\vortT\rVert^2 - \lVert\strainrateT\rVert^2 \right) , 
\end{align}
where the squared norms are $\lVert\tens{A}\rVert^2 = A_{ij}A_{ij}$. 
The $\qcrit\spacetime$ is shown in \movie{mov:Qcrit} and iso-surfaces of $\qcrit=0$ are identified using the skimage python library~\cite{scikit2014}. 

If $\qcrit>0$, vorticity dominates over strain-rate, where as if $\qcrit<0$, strain-rate dominates. 
Where $\qcrit=0$, the magnitudes of vorticity and strain-rate are equal. 
In 2D, isolines of $\qcrit=0$ must form closed loops, and, in 3D, $\qcrit=0$ contours must form closed iso-surfaces enclosing vorticity dominated regions. 
In 2D, there are only two invariants of $\gradVelT$, and the first $\tr{\gradVelT}=\div\velV = 0$ due to incompressibility. 
Therefore, if $\qcrit=0$, all the invariants of $\gradVelT$ are zero, which indicates all the eigenvalues are zero. 
This in turn means the velocity gradient tensor is singular and nilpotent. 
Simple shear flow is an example, as is any viscometric flow in which streamlines are equi-distance apart because there is no stretching or elongation of the fluid\cite{Huilgol1971,Thompson2005}. 
Thus, we refer to $\qcrit=0$ contours as \textit{viscometric surfaces}.

\subsubsection{Comparing zero-isosurfaces $\qcrit$ with vorticity}\label{sctn:vortStrainrateMag}

Throughout this work, we employ the zero-isolines of $\qcrit$-criterion to identify viscometric surfaces as the boundaries between flow structures. 
It is a fact that the magnitude of the symmetric and antisymmetric velocity gradients balance when $\qcrit=0$ to give finite pure shear. 
However, it is also a possibility that the flow has completely vanishing gradients in the velocity. 
To compare these two possibilities of the $\qcrit$-criterion, we explore contours where contributions from $\vortT$ go to zero. 
In any two-dimensional coordinate basis, $\vortT$ only has off diagonal components $\vort_{21}=-\vort_{12}=\vortvec$ in terms of the vorticity pseudo-vector $\vortvecV$. 
Hence, locations where the rotation-rate tensor vanishes can be simply obtained from a sign change in $\vortvec$. 
Obtaining contours where the strain rate vanishes is less direct since the tensorial components vary with coordinate basis as there is no clear handedness to this object and no pseudovector can be constructed in 2D. 
At locations where the vorticity zero-lines and $\qcrit$-criterion zero-lines coincide, the strain-rate contours must be zero by incompressibility. 
Otherwise, these viscometric lines are where there is finite shear from the balance between vorticity and strain-rate. 

\subsubsection{Distribution of vortex characteristics}\label{sctn:distVortex}

Since viscometric surfaces form the boundary encircling vortex-dominated domains from strain-rate-dominated regions, we quantify the distribution of fluid elements composing each vortex domain (where $\qcrit>0$). 
After identifying the list of points defining the $\qcrit=0$ boundary, vortices are identified as the enclosed regions where $\qcrit>0$. 
A clustering algorithm is applied that identifies the constituent vortex points. 
Vortex area distributions are found as the total number of lattice Boltzmann nodes or experimental pixels identified inside a viscometric surface with $\qcrit>0$. 
We calculate the square of the relative shape anisotropy $\shapeani$ of each vortex, written in terms of the first ($I_1$) and second ($I_2$) invariants of the moment of inertia tensor~\cite{Theodorou1985}, in $d$ dimensions as
\begin{align}
    \label{eq:shapeani}
    \shapeani^2 &= 1-\frac{2d}{d-1}\frac{I_2}{I_1^2} .
\end{align}
The enclosed circulation is defined as $\Gamma=\oint_c \velV \cdot d\vec{l} = \oiint \vortvecV\cdot d\vec{S}$. 
We apply the second definition to simplify the handling of non-zero genus vortices. 
The circulation distribution in \fig{fig:vortexstats} is non-dimensionalized to compare the experimental and simulated distributions by dividing by $\Gamma_0 = \vortvec_0 \actlen^2$. 
The vorticity scale, $\vortvec_0$, was identified as the standard deviation of the vorticity distribution. 
We calculate the active length scale as 
\begin{align}
    \label{eq:actLength}
    \actlen &= \defectdens^{-1/2} , 
\end{align} 
where the defect number density $\defectdens=\left( N_{+1/2}+N_{-1/2} \right) / \left( T L^2 \right)$ for $N_{\pm1/2}$ nematic defects, total timesteps $T$ and system size $L$. 

\subsubsection{Splay-bend parameter}\label{sctn:ssb}
The splay-bend parameter~\cite{thijssen2020active,copar2013}
\begin{align}\label{eq:ssb}
    \ssb &= \partial_i \partial_j Q_{ij} 
\end{align}
is used to visualize bend walls as line-like structures with large negative values~\cite{thijssen2020active}. 
To interpret $\ssb$, consider a uniaxial 2D nematic with constant scalar order $S$. 
Under these conditions $\ssb=S\div \left[ \dirV\left(\div\dirV\right) - \dirV\times\left(\curl\dirV\right) \right]$.
The first term inside the divergence is the splay and the second term is the bend, leading to the interpretation that $\ssb$ represents the difference between the divergence of splay and bend. 
Because the active force density is the divergence of the active stress (\eq{eq:activeStress})
\begin{align}
    \label{eq:actForce}
    \actForceV &= -\act\div\nemT,
\end{align}
in active nematics the splay-bend parameter can be understood to be proportional to the divergence of the active force 
\begin{align}
    \ssb &\sim \div \actForceV .
\end{align}

\subsubsection{Defects and viscometric surfaces}\label{sctn:interactions}

To find the probability distribution of nearest distances between defects and viscometric surfaces ($\qcrit=0$ contours), we calculate the distances for all defects to each point on the boundary and included only the smallest distance (to avoid introducing cut-off distances). 
In the case of mirror-symmetric configurations (\fig{fig:Figure_2}c), only the nearest boundary is counted. 

To calculate the distribution of alignment angles shown in \fig{fig:Figure_2}g, we first group $+1/2$ defects to $\qcrit=0$ contours. 
Each defect is associated to a boundary if the defect resides within 2 simulation or experimental sites away from any point of the boundary (experimental sites use the resolution details explained in \sctn{sctn:imaging}). 
The closest point $j$ between the defect and the list of points on the $\qcrit=0$ boundary is found and is the reference point for tangent vector to be calculated from. 
The tangent vector is then calculated using a forward derivative with the next point in the boundary list $j+1$ (with the forward direction matching the $+1/2$ defect's orientation) as
\begin{align}
    \vec{t} &= \frac{\posV(j+1)-\posV(j)}{\abs{\posV(j+1)-\posV(j)}}, 
\end{align}
where $\posV(j)$ is the position vector of the chosen point $j$ on the $\qcrit=0$ boundary. 
We chose the forward derivative over a centered derivative because the flow structure ahead of the defect is more significant to the future dynamics of the defect trajectory. 
The $+1/2$ defect orientation vector is found by \eq{eq:defectAngle}. 
Each alignment angle is therefore found from the angle between these two vectors and presented in the angle interval of $0\si{\degree}$ and $90\si{\degree}$. 
If the defect is associated to the mirror-symmetric $\qcrit=0$ regime, then the same $+1/2$ defect orientation vector is included for the alignment angle calculation with both boundaries.

To investigate the mirror-symmetry breaking, we sort the configurations based on the three classes of defect/flow configurations (\fig{fig:Figure_2}c-e). 
For each $+1/2$ defect, we calculate $\av{\qcrit}_{\Delta \pos^2}$ as the average $\qcrit$-criterion in a small region $\Delta \pos^2$ set to be a rectangle of seven by six lattice units on either side of the mirror-symmetry axis of the defect director field (\ie the defect tail). 
From $\av{\qcrit}_{\Delta \pos^2}$ three different defect populations can be identified: one group where $\av{\qcrit}_{\Delta \pos^2}$ is close to zero (with mirror symmetry), and two where $\av{\qcrit}_{\Delta \pos^2}$ has a large positive or negative value (the mirror-symmetry broken cases). 
To ensemble average over these three regimes, defects are binned into lowest $10\%$ of $\av{\qcrit}_{\Delta \pos^2}$, middle $80\%$ and highest $10\%$. 
This gives \figsi{fig:flowavrdefect}b, \figsi{fig:flowavrdefect}a and \figsi{fig:flowavrdefect}c respectively.
The results are found to be robust to varying the exact binned regions $\Delta\pos^2$. 
 \figsi{fig:flowavrdefect}a shows the ensemble-averaged case of mirror-symmetric defects at the $\qcrit=0$ intersection, with opposite-handed vortices on either side. 
 \figsi{fig:flowavrdefect}b shows the broken-mirror-symmetry state for defects with clockwise-handed $\qcrit>0$ vortex to the right of the defect head and $\qcrit<0$ to the left and vice-versa in  \figsi{fig:flowavrdefect}c. 
 In both cases, the defect aligns parallel with the $\qcrit=0$ tangent. 
 In addition, the local velocity field in \figsi{fig:flowavrdefect}(b-c) on either side of the defect has the same curvature direction as the vortex handedness, while the mirror-symmetric regime has an inversion to the focal point.

\subsection{Modeling}

\subsubsection{Stokesian solitary defect model}\label{sctn:giomiPrediction}
The $\qcrit$-criterion and flow field around $+1/2$ and $-1/2$ topological defects is shown in \fig{fig:Figure_2}b and f. 
The $\qcrit$-criterion field is determined from the velocity field solutions $\velV_{\pm}$ for a single solitary $\pm1/2$ defects first derived by Giomi, \etal~\cite{giomi2014defect}. 
To find the ideal flow field structure around solitary defects of topological charge $k$, they fixed the director field $\dir_{k} = \cos\left(k\phi\right)\hat{x} + \sin\left(k\phi\right)\hat{y}$ and so Beris-Edwards equation (\eq{eq:nem}) does not evolve. 
They then solve the generalized Navier-Stokes equations (\eq{eq:navier_stokes}) for the steady-state ($D_t \velV = \vec{0}$) in the absence of friction and with negligible elastic stresses, leaving the Stokes equation for the velocity field,
\begin{align}
    \visc \laplace \velV - \grad p + \actForceV &= 0
    \label{eq:stokes}
\end{align}
as a response to the active force density $\actForceV$ (\eq{eq:actForce}). 
For an extensile plus half defect ($k=+1/2$), the active force $\actForceV=-\act\hat{x}/2\pos$ is in the direction of the defect $-\hat{x}$ (bend-side) and decays inversely with distance $\pos$ from the defect core~\cite{giomi2014defect}. This rapid decay of the force is the basis for the simple view of defects as point sources of forcing. 
While a bulk active nematic cannot exert a global net force, the conceit of this ideal Stokesian model is that a single solitary defect exists breaking topological charge neutrality, allowing a net force. 
The solution to \eq{eq:stokes} can be written as
\begin{align}
    \velV\left(\posV\right) &= \int dA^\prime \greenT\left(\posV,\posV^\prime\right) \cdot \actForceV\left(\posV^\prime\right)
    \label{eq:activevelocity}
\end{align}
with the two-dimensional Oseen tensor~\cite{giomi2014defect} 
\begin{align}
    \label{eq:green}
    \greenT\left(\posV,\posV^\prime\right) &= \frac{1}{4\pi\visc}\left[ \left(\log \frac{L}{\abs{\posV-\posV^\prime}}-1\right)\unitT + \frac{(\posV-\posV^\prime)\otimes(\posV-\posV^\prime)}{\abs{\posV-\posV^\prime}^2} \right]
\end{align}
in a system bound by the length scale $L$. 
Substituting in the active force for the fixed, solitary defects gives the velocity fields
\begin{align}
    \velV_{+}(r,\phi) &= -\frac{\act}{12\visc} \left[ \left\{3(R-r)+r\cos2\phi\right\} \hat{x} + r\sin2\phi\hat{y} \right] 
    \label{eq:plushalfvel}\\
    \velV_{-}(r,\phi) &= -\frac{\act r}{12\visc R} \left[ \left\{\frac{3}{4}(r-R)\cos2\phi-\frac{R}{5}\cos4\phi\right\}\hat{x} - \left\{(\frac{3}{4}r-R)\sin2\phi+\frac{R}{5}\sin4\phi\right\}\hat{y} \right] ,
    \label{eq:minushalfvel}
\end{align}
where $\act$ is the activity, $\visc$ is the viscosity and $R$ is an integration length scale due to the logarithmic nature of hydrodynamic interactions in 2D (\sctn{sctn:lineForceModel}). 
These are shown in \fig{fig:Figure_2}b and f. 
Taking derivatives of \eqs{eq:plushalfvel}{eq:minushalfvel} produces the velocity gradient tensor $\gradVelT = \grad \otimes \velV$ and the $\qcrit$-criterion field is calculated as the second invariant of $\gradVelT$. 
These calculations and corresponding plots were performed using Mathematica~\cite{Mathematica}. 
For the $+1/2$ defect case, the $\qcrit$-criterion takes the form,
\begin{align}
    &\qcrit_+= \left(\frac{\visc \act}{6}\right)^2 \left( 3\sin^2\phi - 1 \right).
    \label{eq:QCplushalf}
\end{align}
In this construction, the $+1/2$ defect is aligned with a self-propulsion direction in the (negative) $\hat{x}$ direction, with an angle $\theta = \sin^{-1}\left(3^{-1/2}\right) \approx 35.26\si{\degree}$ to the $\mathcal{Q_+}=0$ line. 
This angle is compared in \fig{fig:Figure_2}g against the observed distribution of alignment angles in simulations and experiments, as a benchmark for an isolated $+1/2$ defect. 
The solution for this angle is predicted independent of the activity, viscosity and overall system size. 

\subsubsection{Stokesian straight bend-wall model}\label{sctn:straightwall}

Since our experimental and numerical results indicate that bend walls are coincident with $\qcrit=0$ and crucial to the spontaneous self-constraint, we explore a series of models for the active flows generated by bend walls to identify which features of the bend walls are vital to this interdependence. 
Bend walls are known to arise from the hydrodynamic bend instability~\cite{Ramaswamy2007} leading to arch-like bends of the director field, reminiscent of N\'{e}el walls~\cite{Patelli2019}. 
First consider bend walls modelled as infinitely long, perfectly straight bands of alternating bend and splay given by $\dirV=\cos\theta\hat{x} + \sin\theta\hat{y}$, where the angle $\theta$ is
\begin{align}
    \label{eq:straightwalls}
    \theta(y;\lambda) &= \pi \cos^2\left(\frac{y}{\lambda}\right),
\end{align}
for a modulation length scale $\lambda$. This generates the force density field $\actForceV=\act\frac{\pi}{\lambda} \sin\frac{2y}{\lambda}\left(\cos2\theta\hat{x}+\sin2\theta\hat{y}\right)$. Since the system has translational symmetry in $x$, only derivatives with respect to $y$ exist for the pressure and velocity. Through the incompressibility equation, $\nabla\cdot\velV=0$, the absense of $x-$gradients requires $\frac{\partial \velV_y}{\partial y}=0$. This leads to only one non-zero contribution to the velocity gradient tensor, $\frac{\partial\velV_x}{\partial y}$, which is obtained through integrating the $x-$momentum equation in \eq{eq:stokes}. The velocity gradient tensor is found as,
\begin{align}
    \gradVelT &= -\act\frac{\sin\left(\pi\cos\left[2y/\lambda\right]\right)}{2\eta}
    \begin{pmatrix}
      0 & 1 \\
      0 & 0
    \end{pmatrix}. 
\end{align}
Since this is upper diagonal at all points (representing simple horizontal shear), it is a viscometric flow and $\qcrit=0$ everywhere. 
Thus, a straight, infinitely long series of kink walls is insufficient to explain the colocation of bend walls and $\qcrit=0$. 

\subsubsection{Wavy bend wall model}\label{sctn:wavywalls}

Next we consider if bend wall curvature explains their coincidence with $\qcrit=0$. 
The bend walls from \sctn{sctn:straightwall} are perturbed into infinitely long sinusoidal waves with 
director field once again of the form $\dirV=\cos\theta\hat{x} + \sin\theta\hat{y}$, where the angle $\theta$ but with angle
\begin{align}
    \theta\left(x,y;\lambda\right) &= \pi \cos^2\left(\frac{\pi x}{4 \lambda}\right) + \frac{\pi y}{\lambda} . 
\end{align}
This director field generates a periodic sequence of bend and splay walls that modulate sinusoidally about slices of $\hat{y}$ (\figsi{fig:wavybendwall}a). 
This initial director field is held fixed and the velocity field is numerically evolved to steady state according to \eq{eq:navier_stokes}. 
This simple model of bend walls is sufficient to generate vortices enclosed by viscometric contours (\figsi{fig:wavybendwall}b). 
The vortices reside on the inside regions of the bend and splay walls with greatest curvature, with alternating handedness generated through the coupling between the generated active force and the walls curvature orientation.
Thus, bend wall curvature is essential to the existence of viscometric surfaces in active nematics. 

To further explore the nonlinear coupling between the modulated director field and the active flow field, we then hold the resulting initial steady-state velocity field fixed, but allow the director field to evolve according to \eq{eq:nem}. 
The bend walls proceed to constrict towards a thin kink line, while the strong divergence in the splay wall breaks open, leaving minimised gradients in splay around the bend walls (\figsi{fig:wavybendwall}c). 
This showcases the active instability that drives growth in bend-type deformations --- even with an iterative progression of the two fields, the bend walls drive flows and the flows in turn concentrates and enhances bend~\cite{Ramaswamy2007}. 
Since the centerline of the bend walls coincides with $\qcrit=0$, the backbones themselves are largely undeformed, while the nearby nematic field is rotated towards a parallel alignment with the bend wall centerline, localising the gradients in bend.
Iterating again, demonstrates that the vortices begin to merge and elongate. 
The centerline of the bend wall still resides on $\qcrit=0$

Overall, this model demonstrates that modulation in bend walls is able to split distinct $\qcrit$ regions without needing defects. 
The geometry of the flow generated by the bend has $\qcrit=0$ on the center of the bend wall, which gives bend wall a protection against deforming --- the active flows can constrict but not cleave the bend walls. 
The shape of the bend wall can only be perturbed by the creation of defects.

\subsubsection{Stokesian line force model of bend walls}\label{sctn:lineForceModel}

The hydrodynamic instability constricts infinitely long bend into narrow kink walls and their curvature facilitates closed $\qcrit=0$ contours (\sctn{sctn:wavywalls}). 
However, after the initial onset of active turbulence, finitely long bend walls continue to arise via constriction and be unzipped by $+1/2$ defects (\movie{mov:SimATssb})~\cite{thampi2014instabilities,shankar2019hydrodynamics}. 
The active force density is non-negligible along the bend walls and so we propose a simplified model of finite narrow kink walls being unzipped by $+1/2$ defects modelled as an idealized line of active force. 
As before (\sctn{sctn:giomiPrediction}), elastic stresses are neglected and the Stokes equation (\eq{eq:stokes}) is solved for the velocity field and $\qcrit=0$ contours. 
This is done for four idealized bend wall conformations: \textit{(i)} a point, \textit{(ii)} a straight finite line, \textit{(iii)} a circle and an arc. 
For all extended bend wall conformations, the Stokes equation (\eq{eq:stokes}) is numerically solved. For a straight finite line a $2000\times1000$ grid is used and $1000\times1000$ for the circle and arc. The force lines are constructed by arranging between $500$ to $1000$ point sources in a linear formation, with constant forcing magnitude orientated parallel to the line tangent. Each point on the discretised line contributes to the overall velocity field through the Oseen tensor (\eq{eq:green}).  

\paragraph{\textit{(i)} Point force}
In the far-field limit away from a solitary bend wall, it is effectively a point force. 
From the Oseen tensor (\eq{eq:green}), crudely modelling a bend walls as a point force $\actForce \hat{x}$ at the origin produces a 2D Stokeslet
\begin{align}
    \label{eq:stokeslet}
    \velV\left(\posV\right) &= \frac{\actForce}{4\pi\visc}\left[ \left(\log \frac{L}{\abs{\posV}} - 1\right) \hat{x} + \frac{ x \posV}{\abs{\posV}^2} \right]. 
\end{align}
Taking the gradient results in the $\qcrit$-criterion
\begin{align}
    \qcrit &= -\left(\frac{\actForce}{4\pi\eta\abs{\posV}}\right)^2\cos{2\theta}, 
\end{align}
which predicts an intersection of viscometric surfaces at an angle $\theta=45^\circ$. 

\paragraph{\textit{(ii)} Straight finite line}

In the vicinity of a finite straight-kink wall, the flow field geometry differs from the point-source-like far-field approximation. A horizontal line of constant active force density is constructed using 1000 point sources arranged across 1000 horizontal lattice points, oriented towards the positive $\hat{x}$ direction. The velocity field and zero isolines of the $\qcrit$-criterion (\eq{eq:Qcriterion}) are shown in \fig{fig:Figure_3}c. The $\qcrit$-criterion retains the mirror symmetry observed for the solitary defect or far-field point source, but now differs by constraining the alignment angle $\theta$ towards more acute values. Smaller defect alignment angles in the proximity of bend walls is more consistent with the peak in \fig{fig:Figure_2}g, compared with the angle predicted by the solitary defect model (\sctn{sctn:giomiPrediction}).

\paragraph{\textit{(iii)} Circle and arc}
To provide intuition of the flow geometry generated by bend walls with curvature, we next consider a circular arc force centered on the origin. 
The arc is parameterised in terms of $s$ that can wrap out an angle in a prescribed interval with endpoints lying between $0$ and $2\pi$. 
Positions on the arc are $\posV^\prime(s) = R\left(\cos s\hat{x}+ \sin s\hat{y}\right)$, where $R$ is the radius. 
As before, the force contour is assumed to have constant magnitude $\actForce$ and orient azimuthally. 
By choosing the forces to follow the arc in a clock-wise sense, the force is $\actForceV^\prime(s) = \actForce \left(\sin s\hat{x} - \cos s\hat{y}\right)$. For the arc, numerical integration is performed using 500 point sources distributed between $s=-\pi/2$ and $s=\pi/2$, while the circle uses 1000 point sources between $s=0$ and $s=2\pi$. The results for the velocity field and $\qcrit$-criterion are shown in \fig{fig:Figure_3}d (circle) and e (arc). For both cases, the $\qcrit$-criterion goes to zero when $\abs{\posV}=R$. This result (\fig{fig:Figure_3}e) shows that this simplified model can capture the correlation between bend walls and $\qcrit=0$. Through comparison with the straight finite line case, curvature facilitates the transformation from two vortices to one at the source of the active forcing. Therefore, lines of active force generated by narrow, bend walls, can capture the essential geometry of the flow field if they are \textit{(i)} finite and \textit{(ii)} curved.

\clearpage
\section{Extended data figures}\label{sctn:extendedFigs}

\renewcommand{\figurename}{Extended Data Figure}
\renewcommand\thefigure{\arabic{figure}}  
\setcounter{figure}{0}   

\begin{figure}[h!]
    \centering
    \includegraphics[width=\columnwidth]{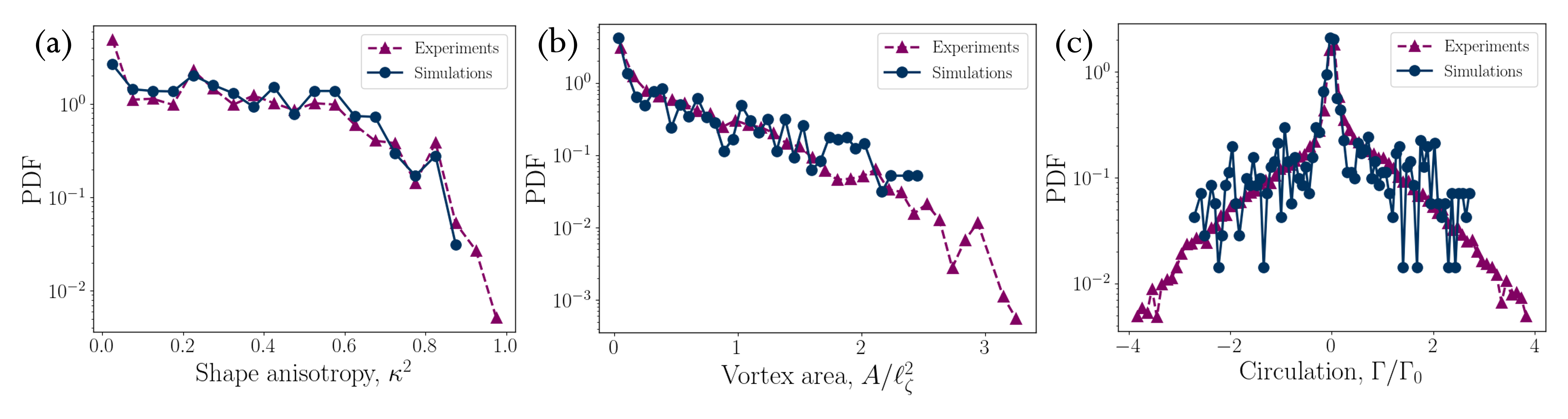}
    \caption{
    \textbf{Probability distribution functions (PDF) of vortex properties, shown for experiments and simulations.} 
    \textbf{(a)} Shape anisotropy parameter $\shapeani^2$ (\eq{eq:shapeani}). 
    The limiting value $\kappa=0$ represents a circle and $\kappa=1$ indicates a line. 
    \textbf{(b)} Vortex area non-dimensionalized by the active length scale $\actlen$. 
    \textbf{(c)} Total circulation within a $\qcrit=0$ closed contour (\sctn{sctn:distVortex}).}
    \label{fig:vortexstats}%
\end{figure}

\begin{figure}[h!]
    \centering
    \includegraphics[width=\columnwidth]{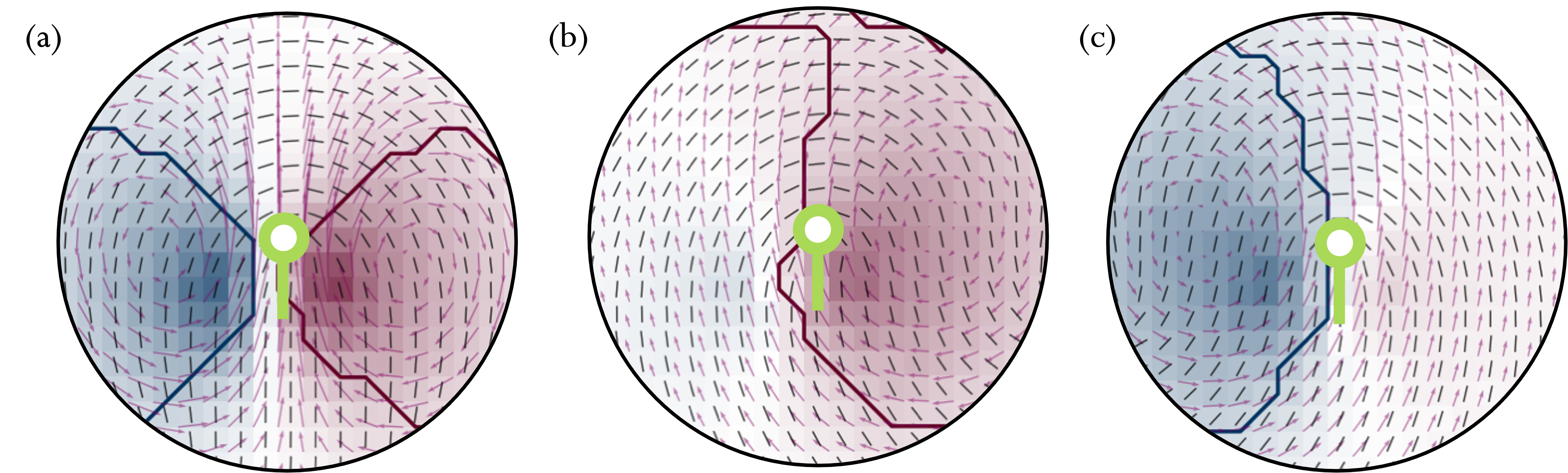}
    \caption{
    \textbf{Time and ensemble averaged director fields and $\qcrit$-criterion around $+1/2$ defects sorted by handedness.} 
    \textbf{(a)} Mirror-symmetric state showing $\qcrit$-criterion structure around $+1/2$ defect. 
    The $\qcrit=0$ contours join towards a point, separating two vortices of opposite handedness. 
    Strain rate dominated flow regions are located in front and behind the defect. 
    \textbf{(b)} Broken-symmetry state with a defect following a clockwise vortex. 
    \textbf{(c)} Same as (b) but following an anti-clockwise vortex. 
    Director field shown as black lines, contours where $\qcrit=0$ are shown as solid lines and colormap corresponds to vorticity field. 
    }
    \label{fig:flowavrdefect}%
\end{figure}

\begin{figure}[h!]
    \centering
    \includegraphics[width=\columnwidth]{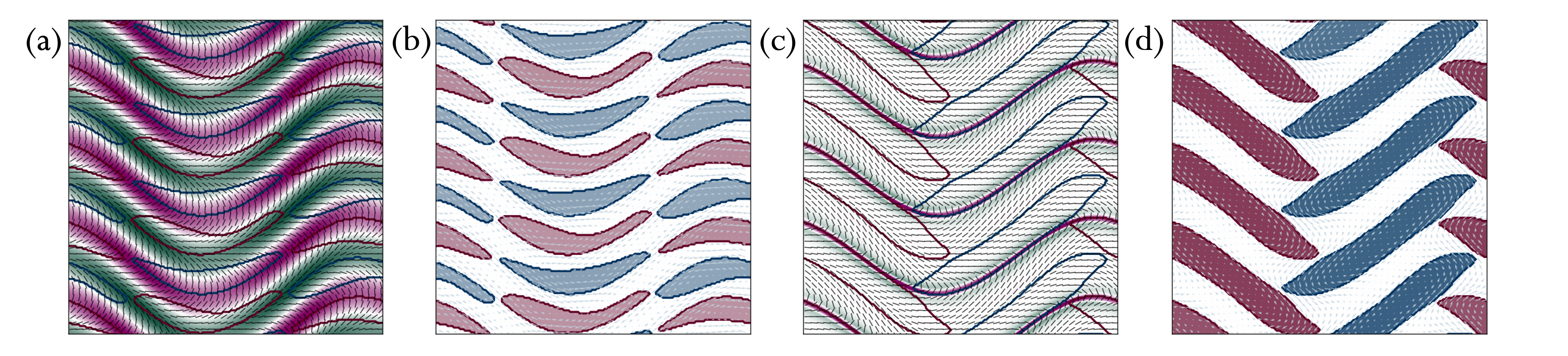}
    \caption{
    \textbf{Infinitely long, repeating wavy bend wall model (\sctn{sctn:wavywalls}).}
    \textbf{(a)} Initial splay-bend parameter ($\ssb$) field with repeating model bend walls in purple, with same colorbar as \fig{fig:Figure_3}a. 
    \textbf{(b)} Resulting $\qcrit$ of steady-state active flows due to fixed $\ssb$ from (a). 
    \textbf{(c)} Resulting steady-state $\ssb$ due to fixed flows from (b). 
    \textbf{(d)} Resulting steady-state flows due to fixed $\ssb$ from (c). Director field is shown as black lines, the velocity field is shown as grey arrows and contours where $\qcrit=0$ are shown as solid lines.
    \label{fig:wavybendwall}%
    }
\end{figure}

\begin{figure}[h!]
    \centering
    \includegraphics[width=\columnwidth]{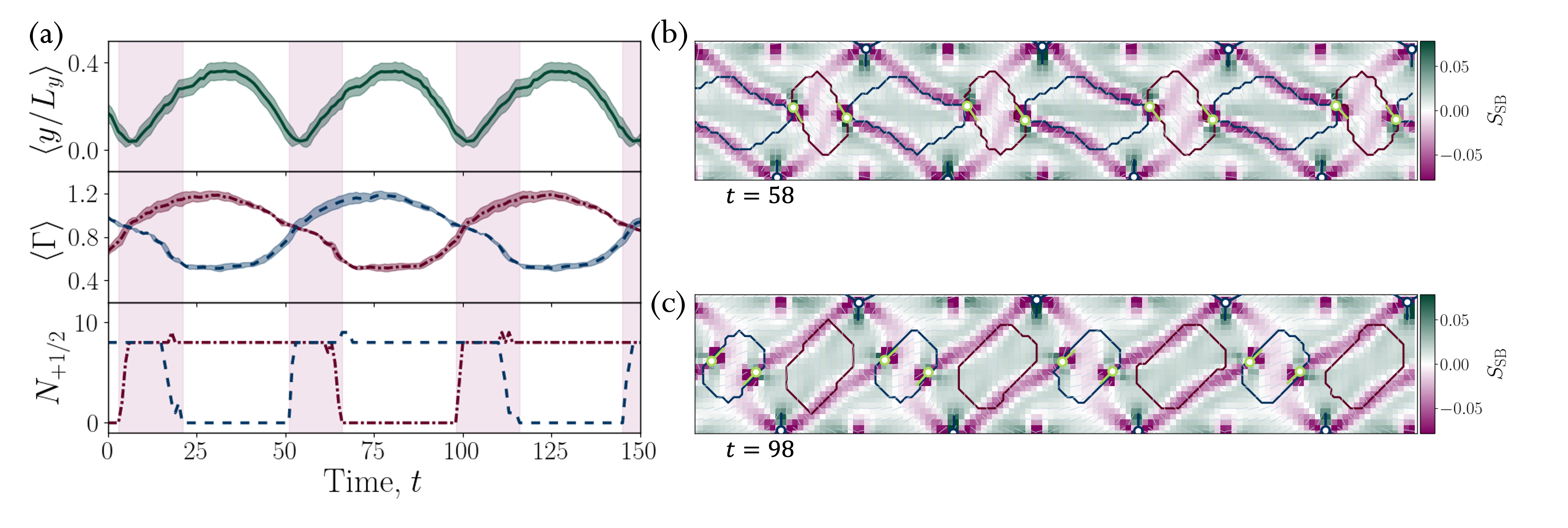}
    \caption{
    \textbf{Numerical simulation of defect dynamics and circulation injection in two-dimensional dance defect state.}
    \textbf{(a)} Defect dynamics and circulation time evolution. 
    The pink shaded region identifies the time periods where at least one defect resides on two vortices of alternate handedness simultaneously. 
    \textbf{(top)} Average absolute +1/2 defect displacement, $y$, from the channel centre, scaled by the channel height $L_y$. 
    Shaded region represents the standard deviation of the defect height.
    \textbf{(middle)} Ensemble average of the circulation, $\Gamma$, contained inside the central clockwise (red) and anti-clockwise (blue) vortices. 
    Shaded region represents the standard deviation for each of the clockwise (red shading) and anti-clockwise (blue shading) vortices. 
    \textbf{(bottom)} Total number of +1/2 defects, $N_{+1/2}$, associated with the central clockwise (red) and anti-clockwise (blue) vortices. 
    \textbf{(b)} Reconnecting bend-wall structure supports the reshaping of viscometric $\qcrit=0$ surfaces. 
    Colormap identifies the line-like nematic bend deformations through the strongly-negative splay-bend parameter $S_{SB}$ (\sctn{sctn:ssb}). 
    Zero-isolines of the $\qcrit$-criterion are shown as solid lines, colored in red for clockwise and blue for anticlockwise. 
    Nematic $+1/2$ defects are marked by green comet-shaped symbols and minus-half defects by dark blue-trefoil shaped symbols (\sctn{sctn:defectMethods}). \textbf{(c)} Same as (b) for a later frame in \movie{mov:ssbVL} when $+1/2$ defects are in their mirror-symetry broken conformation.}
    \label{fig:ceilidh}%
\end{figure}


\clearpage
\section{Supplementary Videos}\label{sctn:movies}


\begin{enumerate}[label=Supplementary Video \arabic*, leftmargin=*]
    \item\label{mov:ExpAT} 
    \textbf{Fluorescence microscopy video of a bundled microtubule/kinesin network at the oil-water interface exhibiting active turbulence.} 
    Solid lines indicate the viscometric surfaces ($\qcrit=0$). 
    Topological $\pm1/2$ defects marked by green comet-like and blue trefoil-like symbols.
    \FIG{fig:Figure_1}a from this video, and scale is the same. 
    %
    \item \label{mov:SimATCirculation} 
    \textbf{Numerically simulated, two-dimensional active turbulence for extensile and flow-aligning nematic (\sctn{sctn:sims}).} 
    Colormap shows the circulation inside vorticity-dominated regions ($\qcrit>0$). 
    Strain-rate-dominated regions ($\qcrit<0$) uncolored. 
    Solid lines indicate the viscometric surfaces ($\qcrit=0$). 
    Topological $\pm1/2$ defects marked by green comet-like and blue trefoil-like symbols. 
    \FIG{fig:Figure_1}b from this video. 
    %
    \item \label{mov:SimATVelGradComp} 
    \textbf{Same as \movie{mov:SimATCirculation} with colormap showing the vorticity field.} 
    Viscometric surfaces and $\pm1/2$ defects illustrated as in \movie{mov:SimATCirculation}. 
    Lines of zero-vorticity contours are shown as orange-dash-dot lines. 
    \FIG{fig:Figure_1}f from this video. 
    %
    \item \label{mov:SimATssb} 
    \textbf{Same as \movie{mov:SimATCirculation} with colormap showing the splay-bend parameter, $\ssb$, field.} 
    Viscometric surfaces and $\pm1/2$ defects illustrated as in \movie{mov:SimATCirculation}. 
    \FIG{fig:Figure_3}a from this video. 
    %
    \item \label{mov:SimFT}
    \textbf{Numerically simulated, two-dimensional active turbulence in the flow tumbling regime (\sctn{sctn:sims}).} 
    Circulation, viscometric surfaces and $\pm1/2$ defects illustrated as in \movie{mov:SimATCirculation}. 
    \FIG{fig:Figure_4}a from this video. 
    %
    \item \label{mov:SimContractile} 
    \textbf{Numerically simulated, two-dimensional active turbulence with contractile activity (\sctn{sctn:sims}).} 
    Splay-bend parameter, viscometric surfaces and $\pm1/2$ defects illustrated as in \movie{mov:SimAniFric}. 
    \FIG{fig:Figure_4}b from this video. 
    %
    \item \label{mov:SimAniFric}
    \textbf{Numerically simulated, two-dimensional active turbulence with anisotropic friction (\sctn{sctn:sims}).} 
    Circulation, viscometric surfaces and $\pm1/2$ defects illustrated as in \movie{mov:SimATCirculation}. 
    \FIG{fig:Figure_4}c from this video. 
    %
    \item \label{mov:SimVLCirc} 
    \textbf{Numerically simulated, two-dimensional dancing defects and vortex lattice in channel with impermeable, no-slip, homogeneous-anchored walls (\sctn{sctn:sims}).} 
    Circulation, viscometric surfaces and $\pm1/2$ defects illustrated as in \movie{mov:SimATCirculation}. 
    \FIG{fig:Figure_4}d-e from this video. 
    %
    \item \label{mov:ExpVL} 
    \textbf{Experimental microscopy of bundled microtubule/kinesin network confined between micro-printed grid walls (\sctn{sctn:expChannel}).} 
    Fluorescence microscopy, viscometric surfaces and $\pm1/2$ defects illustrated as in \movie{mov:ExpAT}. \FIG{fig:Figure_4}f from this video, and scale is the same. 
    %
    \item \label{mov:ssbVL} 
    \textbf{Same as \movie{mov:SimVLCirc} with colormap showing the splay-bend parameter, $\ssb$, field.} 
    Splay-bend parameter, viscometric surfaces and $\pm1/2$ defects illustrated as in \movie{mov:SimAniFric}. \figsi{fig:ceilidh}b-c from this video.
    %
    \item \label{mov:3DVL} 
    \textbf{Numerically simulated, three-dimensional dancing defects and vortext lattice in square duct with impermeable, no-slip, homogeneous-anchored walls (\sctn{sctn:sims}).} 
    Grey shading demarcates vorticity-dominate ($\qcrit>0$) and strain-rate dominate ($\qcrit<0$) space. 
    Disclinations lines visualized as thin tubes colored $\cos\beta$ (\sctn{sctn:defectMethods}), where yellow indicates $\cos\beta\approx+1$ representing local $+1/2$ wedge profiles, green $\cos\beta\approx0$ for twist-type profiles and blue $\cos\beta\approx-1$ for $-1/2$ wedge profiles. 
    \FIG{fig:Figure_4}g from this video. 
    %
    \item \label{mov:Qcrit}
    \textbf{Same as \movie{mov:SimATCirculation} with colormap showing the $\qcrit$-criterion.} 
    Viscometric surfaces and $\pm1/2$ defects illustrated as in \movie{mov:SimATCirculation}.    
\end{enumerate}


\end{document}